\documentclass[%
reprint,
superscriptaddress,
 amsmath,amssymb,
 aps
]{revtex4-1}

\usepackage{graphicx}
\usepackage{dcolumn}
\usepackage{bm}
\usepackage[frozencache]{minted}
\usepackage{mhchem}
\usepackage{todonotes}
\usepackage{multirow}
\usepackage{booktabs}
\usepackage{caption}
\usepackage{siunitx}

\newcommand\norm[1]{\left\lVert#1\right\rVert}

\newenvironment{code}{\captionsetup{type=listing}}{}

\begin{document}


\title{SchNetPack: A Deep Learning Toolbox For Atomistic Systems}

\author{K.T.~Sch\"utt}
\email{kristof.schuett@tu-berlin.de}
\affiliation{Machine Learning Group, Technische Universit\"at Berlin, 10587 Berlin, Germany}
\author{P.~Kessel}
\affiliation{Machine Learning Group, Technische Universit\"at Berlin, 10587 Berlin, Germany}
\author{M.~Gastegger}
\affiliation{Machine Learning Group, Technische Universit\"at Berlin, 10587 Berlin, Germany}
\author{K.~Nicoli}
\affiliation{Machine Learning Group, Technische Universit\"at Berlin, 10587 Berlin, Germany}
\author{A.~Tkatchenko}
\email{alexandre.tkatchenko@uni.lu}
\affiliation{Physics and Materials Science Research Unit, University of Luxembourg, L-1511 Luxembourg, Luxembourg}
\author{K.-R.~M\"uller}
\email{klaus-robert.mueller@tu-berlin.de}
\affiliation{Machine Learning Group, Technische Universit\"at Berlin, 10587 Berlin, Germany}
\affiliation{Department of Brain and Cognitive Engineering, Korea University, Anam-dong, Seongbuk-gu, Seoul 136-713, South Korea}
\affiliation{Max-Planck-Institut f\"ur Informatik, Saarbr\"ucken, Germany}

\begin{abstract}
SchNetPack is a toolbox for the development and application of deep neural networks to the prediction of potential energy surfaces and other quantum-chemical properties of molecules and materials.
It contains basic building blocks of atomistic neural networks, manages their training and provides simple access to common benchmark datasets.
This allows for an easy implementation and evaluation of new models.
For now, SchNetPack includes implementations of (weighted) atom-centered symmetry functions and the deep tensor neural network SchNet as well as ready-to-use scripts that allow to train these models on molecule and material datasets.
Based upon the PyTorch deep learning framework, SchNetPack allows to efficiently apply the neural networks to large datasets with millions of reference calculations as well as parallelize the model across multiple GPUs.
Finally, SchNetPack provides an interface to the \emph{Atomic Simulation Environment} in order to make trained models easily accessible to researchers that are not yet familiar with neural networks.
%
\end{abstract}

\maketitle


\section{\label{sec:introduction}Introduction}

One of the fundamental aims of modern quantum chemistry, condensed matter physics and materials science is to numerically determine the properties of molecules and materials. 
Unfortunately, the computational cost of accurate calculations prove prohibitive when it comes to large-scale molecular dynamics simulations or the exhaustive exploration of the vast chemical space.
Over the last years however, it has become clear that machine learning is able to provide accurate predictions of chemical properties at significantly reduced computational costs. 
Conceptually, this is achieved by training a machine learning model to reproduce the results of reference calculations given the configuration of an atomistic system. 
Once trained, predicting properties of other atomistic systems is generically cheap and has been shown to be sufficiently accurate for a range of applications~\cite{bartok2010gaussian,rupp2012fast,schutt2014represent,behler2015constructing,huo2017unified,faber2018alchemical,de2016comparing,morawietz2016van,gastegger2017machine,chmiela2017machine,faber2017prediction,podryabinkin2017active,brockherde2017bypassing,bartok2017machine,schutt2018schnet,chmiela2018towards,ziletti2018insightful,dragoni2018achieving}.

A common subclass of machine learning models for quantum-chemistry are \emph{atomistic neural networks}.
There exist various architectures of these models, which can be broadly split into two categories: descriptor-based models that take a predefined representation of the atomistic system as input~\cite{behler2007generalized,montavon2012learning,montavon2013machine,zhang2018deep,smith2017ani,gastegger2018wacsf} and end-to-end architectures that learn a representation directly from atom types and positions~\cite{schuett2017quantum,gilmer2017neural,schutt2017schnet,lubbers2018hierarchical}.

SchNetPack provides a unified framework for both categories of neural networks.
While we plan to support more architectures in the future, SchNetPack currently includes implementations for SchNet~\cite{schutt2017schnet,schutt2018schnet}, an end-to-end continuous convolution architecture, as well as Behler--Parrinello networks which are based on atom-centered symmetry functions (ACSF)~\cite{behler2007generalized,behler2011atom} and an extension thereof which uses weighted atom-centered symmetry functions (wACSF)~\cite{gastegger2018wacsf}.

SchNetPack furthermore contains functionality for accessing popular benchmark datasets, training neural networks on (multiple) GPUs to predict a variety of chemical properties. 
It is built in an extensible manner and is implemented using the PyTorch deep learning framework.

The remainder of the paper is structured as follows.
In Section~\ref{sec:model}, we present how models in SchNetPack are structured and briefly review (w)ACSF and SchNet representations.
Section~\ref{sec:training} outlines how SchNetPack manages the training process for atomistic neural networks and gives an overview of the integrated datasets.
Section~\ref{sec:implementation} summarizes details about the implementation, while Sections~\ref{sec:example} and \ref{sec:spkfc} provide code examples for training an atomistic neural network and calculating a power spectrum using the interface to the Atomic Simulation Environment (ASE).
Section~\ref{sec:results} presents results of SchNetPack on standard benchmarks, before we conclude and give an outlook on future extensions.


\begin{center}
\begin{figure*}[t!]
\centering
\includegraphics[width=0.8\textwidth]{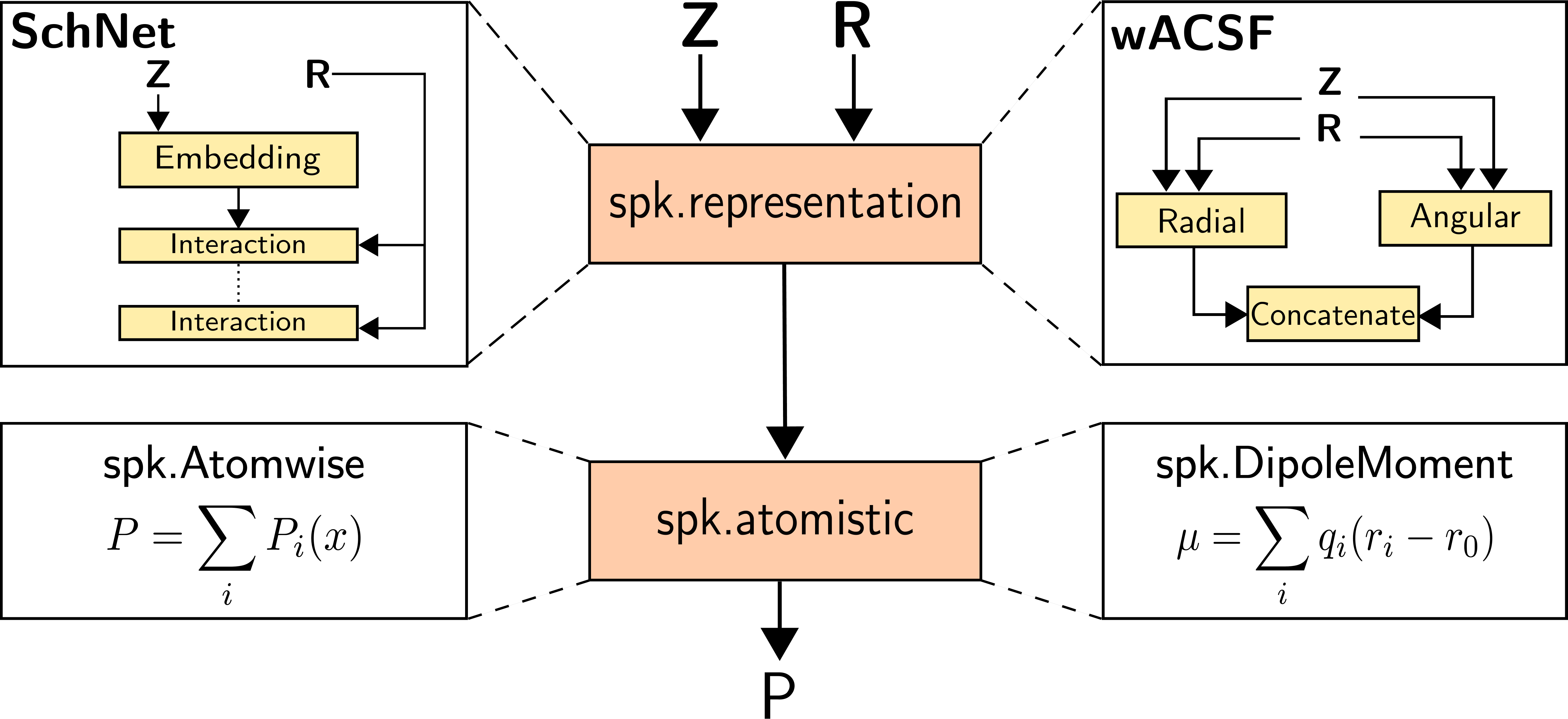}
\caption{Basic building blocks of a model predicting the property $P$ from the positions $R$ and atomic numbers $\mathbf{Z}$ of the atomistic system. We use the abbreviation $\mintinline{python}{spk}$ for the $\mintinline{python}{schnetpack}$ package. All \textit{representation} and \textit{prediction} blocks are collected in the $\mintinline{python}{spk.representation}$ and $\mintinline{python}{spk.atomistic}$ package respectively. The right and left panels illustrate various choices for these building blocks.}
\label{fig:arch}
\end{figure*}
\end{center}


\section{\label{sec:model}Models}
Models in SchNetPack have two principle components: \textit{representation} and \textit{prediction} blocks (see Figure~\ref{fig:arch}).
The former takes the configuration of the atomistic system as an input and generates feature vectors describing each atom in its chemical environment.
The latter uses these atom-wise representations to predict the desired properties of the atomistic system.
The only difference between descriptor-based and end-to-end architectures is whether the representation block is fixed or learned from data.
In the following two sections, we will explain the possible choices for these components in detail.



\subsection{\label{sec:representations}Representations}
An atomistic system containing $n$ atoms can be described by its atomic numbers $\mathbf{Z} = (Z_1, \dots, Z_n)$ and positions $R = (\mathbf{r}_1, \dots, \mathbf{r}_n)$. 
The interatomic distances are given as $r_{ij}=\norm{\mathbf{r}_i-\mathbf{r}_j}$.
In the following, we will briefly describe the currently implemented representations, i.e. (w)ACSF~\cite{gastegger2018wacsf} and SchNet~\cite{schutt2017schnet}.
For further details, refer to the original publications.

\subsubsection{\label{sec:wACSF}(w)ACSF}

Behler--Parrinello network potentials~\cite{behler2007generalized} have proven very useful for systems as diverse as small molecules, metal and molecular clusters, bulk materials, surfaces, water and solid-liquid interfaces (for a recent review, see \cite{behler2017first}). Due to this impressive number of applications, Behler--Parrinello networks are now firmly established as a highly successful neural network architecture for atomistic systems.

For these networks, so-called atom-centered symmetry functions (ACSFs) form the representation of the atomistic system. 
Contrary to the approach taken by SchNet, where features are learned from the data, ACSFs need to be determined before training. 
Hence, using symmetry functions can be advantageous in situations where the available training data is insufficient to learn suitable representations in an end-to-end fashion.
On the other hand, introducing rigid hand-crafted features might reduce the generality of the model. 
In the following, we will briefly review ACSFs and a variant called weighted ACSFs, or wACSFs for short. 
We refer to References~\cite{behler2011atom} and \cite{gastegger2018wacsf} for a more detailed discussion.

ACSFs describe the local chemical environment around a central atom via a combination of radial and angular distribution functions.

\paragraph{Radial Symmetry Functions:}
Radial ACSF descriptors take the form:
\begin{align}
G_{i,\alpha}^{rad} = \sum_{j \neq i}^{N} g(Z_j) e^{-\gamma_\alpha (r_{ij} -\mu_\alpha)^2} f(r_{ij}), \label{eq:radialWACSF}
\end{align}
where $i$ is the central atom and the sum runs over all neighboring atoms $j$. $\gamma_\alpha$ and $\mu_\alpha$ are parameters which modulate the widths and centers of the Gaussians.
Typically, a set of $n_\mathrm{rad}$ radial symmetry functions with different parameter combinations $\alpha \in \{1,\dots,n_\mathrm{rad}\}$ are used.
In SchNetPack, suitable $\gamma_\alpha$ and $\mu_\alpha$ are determined automatically via an equidistant grid between zero and a spacial cutoff $r_c$, adopting the empirical parametrization strategy detailed in Reference~\cite{gastegger2018wacsf}.

A cutoff function $f$ ensures that only atoms close to the central atom $i$ enter the sum and is given by
\begin{align}
f(r_{ij}) = \begin{cases}
\tfrac12 \left( \cos(\tfrac{\pi r_{ij}}{r_c}) + 1 \right) \,, & r_{ij} \leq r_c \,, \\
0 \,, & \, \text{else} \,.
\end{cases}
\end{align}
For convenience, we will use the notation $f_{ij}=f(r_{ij})$ in the following. 
Finally, $g(Z_j)$ is an element-dependent weighting function. In ACSFs, $g(Z_j)$ takes the form
\begin{align}
g(Z_j)=\delta_{Z_j,Z_\mathfrak{a}} =  \begin{cases} 1 & \mbox{if } Z_j=Z_\mathfrak{a} \\  0 & \mbox{else}. \end{cases}
\end{align}
Hence, radial ACSFs are always defined between the central atom and a neighbor belonging to a specific chemical element.

\paragraph{Angular Symmetry Functions:}
information about the angles between atoms are encoded by the $n_a$ angular symmetry functions
\begin{align}
	G_{i,\alpha}^{ang}=&2^{1-\zeta_\alpha} \sum^N_{j\neq i,k > j} g(Z_j, Z_k) \left( 1 + \lambda \theta_{ijk} \right)^{\zeta_\alpha} \, \nonumber \\
	& \times e^{-\gamma_\alpha \left( r_{ij}^2 + r_{ik}^2 + r_{jk}^2 \right)}   f_{ij} \, f_{ik} \, f_{jk} \,, \label{eq:angularWACSF}
\end{align} 
where $\theta_{ijk}$ is the angle spanned between atoms $i$, $j$ and $k$. The parameter $\lambda$ takes the values $\lambda=\pm 1$ which shifts the maximum of the angular terms between $0$ and $\pi$. The variable $\zeta_\alpha$ is a hyperparameter controlling the width around this maximum. $\gamma_\alpha$ once again controls the width of the Gaussian functions. As with radial ACSFs, a set of $n_\mathrm{ang}$ angular functions differing in their parametrization patterns $\alpha \in \{1,\dots,n_\mathrm{ang}\}$ is chosen to describe the local environment.
For angular ACSFs, the weighting function $g(Z_k, Z_j)$ can be expressed as
\begin{align}
g(Z_k,Z_j) = \frac{1}{2}\left( \delta_{Z_j Z_\mathfrak{a}} \delta_{Z_k Z_\mathfrak{b}} +  \delta_{Z_j Z_\mathfrak{b}} \delta_{Z_k Z_\mathfrak{a}} \right),
\end{align}
which counts the contributions of neighboring atoms $j$ and $k$ belonging to a specific pair of elements (e.g. O-H or O-O).

Due to the choice of $g$, ACSFs always are defined for pairs (radial) or triples (angular) of elements and at least one parametrized function $G_\mathrm{i,\alpha}$ has to be provided for each of these combinations. As a consequence, the number of ACSFs grows quadratically with the number of different chemical species. This can lead to an impractical number of ACSFs for systems containing more than four elements (e.g. QM9).

Recently, alternative weighting functions have been proposed which circumvent the above issue. In these so-called weighted ACSFs (wACSFs), the radial weighting function is chosen as $g(Z_j)=Z_j$ while the angular function is set to $g(Z_k,Z_j)=Z_k Z_j$. Through this simple reparametrization, the number of required symmetry functions becomes independent of the actual number of elements present in the system, leading to more compact descriptors. SchNetPack uses wACSFs as the standard descriptor for Behler--Parrinello potentials.

Irrespective of the choice for the weighing $g$, both radial and angular symmetry functions are concatenated as a final step to form the representation for the atomistic system, i.e.
\begin{align}
    X_i = \left( G^{rad}_{i,1}, \dots G^{rad}_{i,n_\mathrm{rad}}, G^{ang}_{i,1}|_{\lambda=\pm 1}, \dots, , G^{ang}_{i,n_\mathrm{ang}}|_{\lambda=\pm 1} \right) \,. \label{eq:featureswACSF}
\end{align}
This representation $X_i$ can then serve as input for prediction block of the atomistic network.

\subsubsection{\label{sec:schnet}SchNet}
SchNet is an end-to-end deep neural network architecture based on continuous-filter convolutions~\cite{schutt2017schnet, schutt2018schnet}.
It follows the deep tensor neural network framework~\cite{schuett2017quantum}, i.e. atom-wise representations are constructed by starting from embedding vectors that characterize the atom type before introducing the configuration of the system by a series of interaction blocks.

Convolutional layers in deep learning usually act on discretized signals such as images. 
Continuous-filter convolutions are a generalization thereof for input signals that are not aligned on a grid, such as atoms at arbitrary positions. 
Contrary to (w)ACSF networks which are based on rigid hand-crafted features, SchNet adapts the representation of the atomistic system to the training data.
More precisely, SchNet is a multi-layer neural network which consists of an embedding layer and several interaction blocks, as shown in the top left panel of Figure~\ref{fig:arch}.
We describe its components in more detail in the following:

\paragraph{Atom Embeddings:} 
Using an embedding layer, each atom type $Z_i$ is represented by feature vectors $\mathbf{x}^0_i\in \mathbb{R}^F$ which we collect in a matrix $X^0 = (\mathbf{x}^0_1, \dots ,\mathbf{x}^0_n)$. 
The feature dimension is denoted by $F$. 
The embedding layer is initialized randomly and adapted during training. 
In all other layers of SchNet, atoms are described analogously and we denote the features of layer $l$ by $X^l = (\mathbf{x}^l_1, \dots, \mathbf{x}^l_n)$ with $\mathbf{x}^l_i\in \mathbb{R}^F$.

\paragraph{Interaction Blocks:}
Using the features $X^l$ and positions $R$, this building block computes interactions which additively refine the previous representation analogue to ResNet blocks~\cite{he2016deep}.
To incorporate the influence of neighboring atoms, continuous-filter convolutions are applied which are defined as follows:
\begin{equation}
\mathbf{x}_i^{l+1} = (X^l * W^l) \equiv \sum_{j \in \text{nbh}(i)} \mathbf{x}_j^l \odot W^l(r_{ij}) \,.
\end{equation}
By $\odot$ we denote element-wise multiplication and $\text{nbh}(i)$ are the neighbors of atom $i$.
In particular for larger systems, it is recommended to introduce a radial cutoff.
For our experiments, we use a distance cutoff of 5{\AA}.

Here, the filter is not a parameter tensor as in standard convolutional layers, but a filter-generating neural network $W^l: \mathbb{R} \to \mathbb{R}^F$ which maps atomic distances to filter values.
The filter generator takes atom positions expanded on a grid of radial basis functions which are closely related to the radial symmetry functions \eqref{eq:radialWACSF} of (w)ACSF. For its precise architecture, we refer to the original publications~\cite{schutt2017schnet, schutt2018schnet}. 

Several atom-wise layers, i.e. fully-connected layers
\begin{equation} 
\mathbf{x}^{l+1}_i = W^l \mathbf{x}^l_i + \mathbf{b}^l \, \label{eq:atomwise}
\end{equation}
 that are applied to each atom $i$ separately, recombine the features within each atom representation.
Note that the weights $W^l$ and biases $\mathbf{b}^l$ are independent of $i$ and are therefore the same for all atom features $\mathbf{x}^l_i$. 
Thus the number of parameters of atom-wise layers is independent of the number of atoms $n$.

In summary, SchNet obtains a latent representation of the atomistic system by first using an embedding layer to obtain features $X^0$. These features are then processed by $L$ interaction blocks which results in the latent representation $X^L$ which can be passed to the prediction block. We will sketch the possibilities for the architectures of these prediction blocks in the following section.

\subsection{\label{sec:output}Prediction Blocks}
As discussed in the last sections, both SchNet and (w)ACSF provide representations $X_i$ with $i\in\{1,\dots,n\}$ for an atomistic system with $n$ atoms. These representations are then processed by a prediction block to obtain the desired properties of the atomistic system.
There are various choices for prediction blocks depending on the property of interest.
Usually, prediction blocks consist of several atom-wise layers \eqref{eq:atomwise} with non-linearities, which reduce the feature dimension, followed by a property-dependent aggregation across atoms. 

The most common choice are \mintinline{python}{Atomwise} prediction blocks, which express a desired molecular property $P$ as a sum of atom-wise contributions
\begin{align}
    P = \sum_{i=1}^n p(\mathbf{x}_i) \,.
\end{align}
While this is a suitable model for extensive properties such as the energy, intensive properties, which do not grow with the number of atoms $n$ of the atomistic system, are instead expressed as the average over contributions.

\mintinline{python}{Atomwise} prediction blocks are suitable for many properties, however property-specific prediction blocks may be used to incorporate prior knowledge into the model.
The \mintinline{python}{DipoleMoment} prediction block expresses the dipole moment $\mathbf{\mu}$ as
\begin{equation}
    \mathbf{\mu} = \sum_{i=1}^n q(\mathbf{x}_i) (\mathbf{r}_i-\mathbf{r}_0) \,,
\end{equation}
where $q: \mathbb{R}^F \to \mathbb{R}$ can be interpreted as latent atomic charges and $\mathbf{r}_0$ denotes the center of mass of the system. 

The \mintinline{python}{ElementalAtomwise} prediction block is different from \mintinline{python}{Atomwise} in that instead of applying the same network to all the atom features $X_i$, it uses separate networks for different chemical elements.
This is particularly useful for (w)ACSF representations. 
Analogously, \mintinline{python}{ElementalDipoleMoment} is defined for the dipole moment.

\section{Data Pipeline and Training}\label{sec:training}
\begin{figure}[h]
\centering
\includegraphics[width=0.5\textwidth]{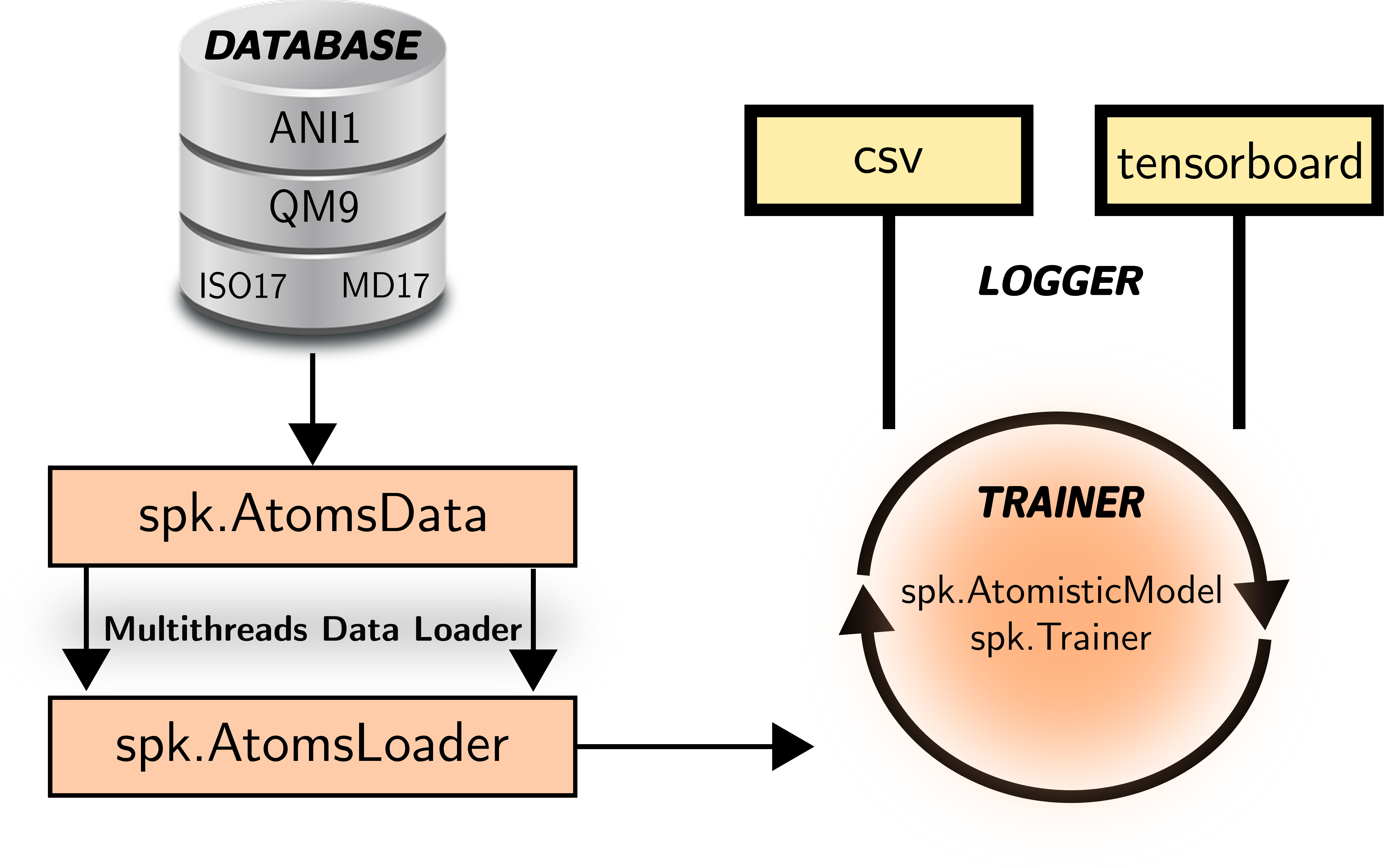}
\caption{Setup for training models in SchNetPack. Note that we denote the various choices for dataset classes mentioned in Section~\ref{sec:training} by their common  \mintinline{python}{spk.AtomsData} base class.}
\label{fig:training}
\end{figure}
One of the main aims of SchNetPack is to accelerate the development and application of atomistic neural networks. To this end, SchNetPack contains a number of classes which provide access to standard benchmark datasets and manage the training process. Figure \ref{fig:training} summarizes this.

The dataset classes automatically download the relevant data, if not already present on disk, and use the standard ASE package~\cite{ase} to store them in an SQLite database. In particular, this means that we use the conventions and units of the ASE package in SchNetPack, e.g. energies and lengths are in units of $\si{\eV}$ and $\si{\angstrom}$. Currently, SchNetPack includes the following dataset classes:
\begin{itemize}
	\item \mintinline{python}{schnetpack.datasets.QM9}: class for the QM9 dataset~\cite{qm9one,qm9two} for 133,885 organic molecules with up to nine heavy atoms from $\ce{C}$, $\ce{O}$, $\ce{N}$ and $\ce{F}$.
	\item \mintinline{python}{schnetpack.datasets.ANI1}: functionality to access ANI-1 dataset~\cite{ani1} which consists of more than 20 million conformations for 57454 small organic molecules from $\ce{C}$, $\ce{O}$ and $\ce{N}$.
	\item \mintinline{python}{schnetpack.datasets.ISO17}: class for ISO17 dataset~\cite{qm9two, schutt2017schnet, schuett2017quantum} for molecular dynamics of $\ce{C_7O_2H_{10}}$ isomers. It contains 129 isomers with 5000 conformational geometries and their corresponding energies and forces.
	\item \mintinline{python}{schnetpack.datasets.MD17}: class for MD17 dataset~\cite{chmiela2017machine, schuett2017quantum} for molecular dynamics of small molecules containing molecular forces.
	\item \mintinline{python}{schnetpack.datasets.MaterialsProject}: provides access to the Materials Project~\cite{mp1} repository of bulk crystal containing atom types ranging across the whole periodic table up to $Z = 94$.
\end{itemize}
We also provide a \mintinline{python}{AtomsLoader} class for feeding a model with (a subset of) a dataset during training using multiple threads. 
This class also calculates relevant statistics such as mean and standard deviation.  

For convenience, a \mintinline{python}{Trainer} class is included in SchNetPack which manages the training process of the model. 
This class evaluates the model's performance on a validation set, provides functionality for early stopping and various learning rate schedules as well as checkpointing and logging. 
For the latter, one can choose between csv files and Tensorboard~\cite{tensorflow} which is a powerful web-based visualization interface. SchNetPack supports training on multiple GPUs for which we use the standard PyTorch implementation.

As we will show in the example discussed in Section~\ref{sec:example}, the classes presented in this section allow us to efficiently train atomistic neural networks and evaluate their performance using a very compact amount of code. 

\section{\label{sec:implementation}Implementation Details}
SchNetPack is implemented in Python using the PyTorch ($\geq$0.4) deep learning library~\cite{pytorch}.
Calculations that do not require automatic differentiation are performed using Numpy~\cite{numpy}.
SchNetPack is tightly integrated with the Atomic Simulation Environment (ASE)~\cite{ase} which is used to persist configurations of atomistic systems. 
We also provide an interface to the ASE calculator class which allows to easily incorporate SchNetPack models into ASE workflows, such as performing molecular dynamics.
Logging the training progress to Tensorboard~\cite{tensorflow} is facilitated by tensorboardX~\cite{tensorboardX}. Some of the datasets come in the HDF5 binary file format which we parse with the h5py package~\cite{h5py}. SchNetPack can be easily installed using pip \footnote{To install run the following command: pip install schnetpack}. The code for SchNetPack can be found on GitHub \footnote{Code can be found here: https://github.com/atomistic-machine-learning/schnetpack}.

\section{\label{sec:example}Example: Training in SchNetPack}

\begin{code}
\begin{minted}[python3=true, xleftmargin=15pt, numbersep=7pt, numbers=left, mathescape]{python}
import schnetpack as spk
import schnetpack.atomistic as atm
import schnetpack.representation as rep
import torch
from torch.optim import Adam
import torch.nn.functional as F
from schnetpack.datasets import *

# load qm9 dataset and download if necessary $\label{l:data}$
data = QM9("qm9/", properties=[QM9.U0])

# split in train and val
train, val, test = data.create_splits(10000,
                                        1000)
loader = spk.AtomsLoader(train, 
                         batch_size=100,
                         num_workers=4)
val_loader = spk.AtomsLoader(val)

# create model $\label{l:rep}$
reps = rep.SchNet()
output = atm.Atomwise()
model = atm.AtomisticModel(reps, output)

# create trainer
opt = Adam(model.parameters(), lr=1e-4)
loss = lambda b,p: F.mse_loss(p["y"],b[QM9.U0])
trainer = spk.Trainer("output/", model, loss,
                      opt, loader, val_loader)

# start training
trainer.train(torch.device("cpu"))
\end{minted}
\caption{Minimal code example for training a SchNet model on the QM9 dataset with SchNetPack.\label{listing:qm9}}
\end{code}

Listing \ref{listing:qm9} is a minimal example of how to train a model with SchNet representation to predict the total energy $U_0$ on QM9. 
Training and validation sets with 10k and 1k datapoints are used and the data is loaded asynchronously using four worker threads.

In order to train on a different dataset, one has to only change line~10 in Listing \ref{listing:qm9}. 
In the example of ANI-1, it will read \mint{python}|data = ANI1("ani1/", properties=[ANI1.energy])|
Similarly, one can straightforwardly change the representation to wACSF by replacing line 21 by
\mint{python}|reps = rep.BehlerSFBlock()|
In this case however, it is advantageous to use a \mintinline{python}{ElementalAtomwise} output network by changing line 22 to
\mint[]{python}|output = atm.ElementalAtomwise(reps.n_symfuncs)|
These examples can also be found in the SchNetPack source directory in the examples subdirectory.

\section{\label{sec:spkfc}Example: SchNetPack for Chemists}

In addition to the above features, SchNetPack provides an interface to the ASE \mintinline{python}{Calculator} class. This makes it possible to use SchNetPack models with the calculation tools available in ASE, such as geometry optimization, normal mode analysis and molecular dynamics simulations.

\begin{figure}[ht]
	\centering
	\includegraphics[width=0.45\textwidth]{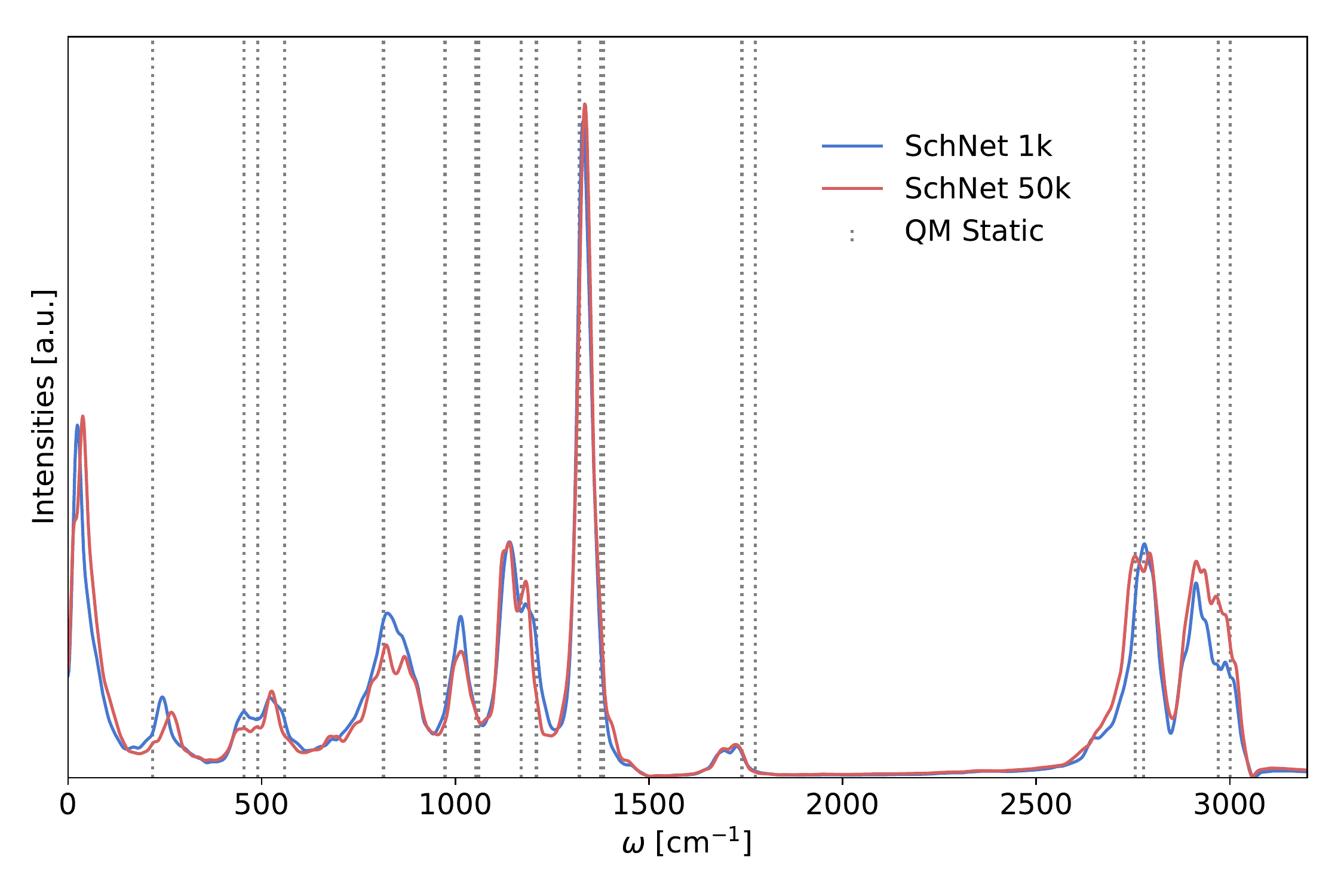}
	\caption{Power spectra of malondialdehyde at 300~K, using SchNets trained on 1000 and 50000 data points taken from the MD17 database. The harmonic normal mode vibrations obtained with the electronic structure reference are shown in grey.}
	\label{fig:spectrum}
\end{figure}

The ASE interface is provided via the \mintinline{python}{AseDriver} class in the \mintinline{python}{molecular_dynamics} module.
\begin{code}
\begin{minted}[python3=true, xleftmargin=15pt, numbersep=7pt, numbers=left, mathescape]{python}

import schnetpack.molecular_dynamics as md

# Load trained model
model = md.load_model(model_directory)

ml_calculator = md.AseInterface( 
    path_to_molecule, model, 
    simulation_directory
    )

# Optimize structure
ml_calculator.optimize()

# Compute numerical normal models
ml_calculator.compute_normal_modes()

# Setup and run molecular dynamics
ml_calculator.init_md()
ml_calculator.run_md()

\end{minted}
\caption{Minimal code example for performing ASE calculations with a trained SchNetPack model stored in \mintinline{python}{model_directory}. \label{listing:ase}}
\end{code}
Listing~\ref{listing:ase} shows an example on how trained models are loaded into the calculator and used for computation.
For convenience, SchNetPack provides the script \mintinline{python}{schnetpack_molecular_dynamics.py} which can be used to perform various simulations out of the box.
To demonstrate the above features, SchNetPack was used to predict the power spectrum of the keto form of malondialdehyde via molecular dynamics simulations (shown in Figure~\ref{fig:spectrum}).
The machine learning models are able to reproduce the peak positions accurately, even when trained on the smaller data set, demonstrating the efficacy of the force training procedure. Particularly impressive are the fine details observed in the spectra. For example, the two models are able to resolve the structure of the peak at 1700~cm$^{-1}$ which is due to the symmetric and asymmetric stretching vibrations of the two carbonyl groups.

SchNet simulations of malondialdehyde take approximately 11~milliseconds per timestep on a Tesla P100 GPU. This corresponds to a speedup of almost three orders of magnitude compared to the original electronic structure reference computations.
In the present setup, Behler--Parrinello networks show a comparable performance to SchNet which indicates that both models do not yet exhaust the full capacity of the GPU for molecules of this size. It can be expected that ACSF based models are more efficient when simulating larger systems and also when using CPUs instead of GPUs.

\section{Results}\label{sec:results}
\begin{table*}
\caption{Summary of performance on test set. By $N$, we denote the size of the combined train and validation set. \label{tab:results}}
\centering
\begin{ruledtabular}
\begin{tabular}{llllrrr} 
Dataset & Property & Unit & Model & MAE & RMSE & time \\
\colrule
\multirow{6}{*}{Malondialdehyde (N=1k)} &  \multirow{3}{*}{energy} & \multirow{3}{*}{kcal mol$^{-1}$} & SchNet & 0.08 & 0.11 & 2.5h \\ 
& & &  ACSF & 0.30 & 0.40 & 0.6h \\ 
& & & wACSF & 1.16 & 1.52 & 0.6h \\ 
&  \multirow{3}{*}{atomic forces} & \multirow{3}{*}{kcal mol$^{-1}$ {\AA}$^{-1}$} & SchNet & 0.13 & 0.16 & 2.5h \\ 
& & &  ACSF & 1.08 & 1.59 & 0.6h \\ 
& & & wACSF & 3.27 & 4.53 & 0.6h \\ 
\colrule
\multirow{6}{*}{Malondialdehyde (N=50k)} &  \multirow{3}{*}{energy} & \multirow{3}{*}{kcal mol$^{-1}$} & SchNet & 0.07 & 0.09 & 13.5h \\ 
& & &  ACSF & 0.09 & 0.11 & 6h \\ 
& & & wACSF & 0.69 & 0.88 & 6h \\ 
&  \multirow{3}{*}{atomic forces} & \multirow{3}{*}{kcal mol$^{-1}$ {\AA}$^{-1}$} & SchNet & 0.05 & 0.09 & 13.5h \\ 
& & &  ACSF & 0.26 & 0.42 & 6h \\ 
& & & wACSF & 1.84 & 2.51 & 6h \\ 
\colrule
\multirow{6}{*}{Acetylsalicylic acid (N=1k)} &  \multirow{3}{*}{energy} & \multirow{3}{*}{kcal mol$^{-1}$} & SchNet & 0.38 & 0.52 & 2.5h \\ 
& & &  ACSF & 0.79 & 1.03 & 0.7h \\
& & & wACSF & 2.11 & 2.69 & 0.7h \\ 
&  \multirow{3}{*}{atomic forces} & \multirow{3}{*}{kcal mol$^{-1}$ {\AA}$^{-1}$} & SchNet & 1.17 & 1.68 & 2.5h \\ 
& & &  ACSF & 1.92 & 2.75 & 0.7h \\ 
& & & wACSF & 4.80 & 6.81 & 0.7h \\ 
\colrule
\multirow{6}{*}{Acetylsalicylic acid (N=50k)} &  \multirow{3}{*}{energy} & \multirow{3}{*}{kcal mol$^{-1}$} & SchNet & 0.11 & 0.14 & 2d 11.5h \\ 
& & &  ACSF & 0.40 & 0.53 & 1d 6h \\ 
& & & wACSF & 1.20 & 2.69 & 1d 6h \\ 
&  \multirow{3}{*}{atomic forces} & \multirow{3}{*}{kcal mol$^{-1}$ {\AA}$^{-1}$} & SchNet & 0.14 & 0.19 & 2d 11.5h \\ 
& & &  ACSF & 0.88 & 1.26 & 1d 6h \\
& & & wACSF & 2.31 & 3.14 & 1d 6h \\ 
\colrule
\multirow{6}{*}{QM9 (N=110k)} & \multirow{3}{*}{$U_0$}  & \multirow{3}{*}{kcal mol$^{-1}$} & SchNet & 0.26 & 0.54 & 12h \\
& & &  ACSF & 0.49 & 0.92 & 8h \\ 
& & & wACSF & 0.43 & 0.81 & 6h \\ 
 & \multirow{3}{*}{dipole moment}  & \multirow{3}{*}{Debye} & SchNet & 0.020 & 0.038 & 13h \\
& & &  ACSF & 0.064 & 0.100 & 8h \\ 
& & & wACSF & 0.064 & 0.095 & 8h \\ 
\colrule
\multirow{1}{*}{ANI-1 (N=10.1M)} & \multirow{1}{*}{energy} & \multirow{1}{*}{kcal mol$^{-1}$} & SchNet & 0.55
 & 0.89 & 9d 7h\footnote{We used four Tesla P100 GPUs for data-parallel training.} \\ 
 \multirow{1}{*}{ANI-1 (N=19.8M)} & \multirow{1}{*}{energy} & \multirow{1}{*}{kcal mol$^{-1}$} & SchNet & 0.47
 & 0.77 & 12d 15h\footnote{We used two Tesla P100 GPUs for data-parallel training.} \\ 
\colrule
Materials Project (N=62k) & formation energy & eV / atom & SchNet & 0.041 & 0.088 & 1d 14h \\ 
\end{tabular}
\end{ruledtabular}
\end{table*}

In this section, we present results on QM9, ANI-1, MD17 and Material Project datasets obtained with SchNetPack. 
A summary of the test set performance of both Behler--Parrinello (ACSF and wACSF) and SchNet models can be found in Table~\ref{tab:results}. The reported results are the average of three models trained on different splits of the same size. The Python scripts with which we obtained these results using a Tesla-P100 GPU can be found in the scripts subdirectory of SchNetPack. We refer to Appendix~\ref{app:experiments} for further details on the experiments.

Although Behler--Parrinello networks produce reliable results for a wide range of experiments, they are consistently outperformed by the SchNet architecture. Due to its end-to-end nature, SchNet is able to infer efficient molecular representations in a data driven fashion which leads to an improved flexibility compared to the rigid handcrafted features used in Behler--Parrinello potentials (ACSF and wACSFs). The expressive power of SchNet models is enhanced further by their deep architecture, compared to the shallow atomistic networks used in Behler--Parrinello models.
These features are also advantageous for learning molecular forces for which derivatives of the energy prediction are required for training.
A good example are the results obtained for the molecules malonaldehyde and acetylsalicylic acid taken from the MD17 dataset.
Here, SchNet outperforms the other models even on small training sets.
SchNet achieves chemically accurate performance for data sets containing a wealth of different molecular configurations (ANI-1),
as well as for compounds incorporating a wide range of chemical elements, demonstrating its high utility.

The prime advantage of Behler--Parrinello models is their reduced computational cost compared to SchNet, which is expected to be beneficial e.g. for molecular dynamics simulations of large molecules. Moreover, it should be noted that all ACSF and wACSF models presented here use the empirical scheme introduced in Reference~\cite{gastegger2018wacsf}. Their performance can be improved by careful fine-tuning of the descriptors. However, such a procedure is typically tedious, especially considering the excellent out of the box performance of SchNet.

An interesting effect can be observed when comparing the performance of standard ACSFs to the recently suggested wACSFs.
In tasks which focus on modeling structurally and chemically diverse datasets (QM9), wACSF produce better results.
However in problems for which small variations of the molecular structure need to be resolved (MD17), ACSFs outperform wACSFs. 
The reason for this behavior is the loss of spatial resolution of wACSF which is a direct consequence of the improved elemental resolution. Whether this problem can be circumvented by learning elemental weights in a similar manner as in SchNet will be the focus of future research.

SchNet achieves chemically accurate prediction on the ANI-1 dataset. The ANI-1 neural network potential~\cite{smith2017ani}, which is based on Behler--Parrinello networks, reported a RMSE of 1.2 kcal mol$^{-1}$ using 80\% of the ANI-1 dataset for training and 10\% for validation. Using SchNet, we already obtain a RMSE of 0.89 kcal mol$^{-1}$ using a training set of 10 million reference examples. Raising our splits up to 80\% of the whole dataset for training and 10\% for validation and testing, we obtain a MAE of 0.47 kcal mol$^{-1}$ and a RMSE of 0.77 kcal mol$^{-1}$.

\section{\label{sec:conclusions}Conclusions}

SchNetPack is a framework for neural networks of atomistic systems which simplifies accessing standard benchmark datasets, training models of different architectures and evaluating their performance. It provides an interface to combine it with the functionality of the ASE package such as molecular dynamics simulations.
We plan on extending SchNetPack further in the future by adding more datasets, advanced training mechanisms such as active sampling, support for additional quantum-mechanical observables and further neural network architectures.
We expect this unification and simplification to be of great value for the community as it allows to concentrate on the design of the neural network models as well as to easily compare different architectures.

\section*{Acknowledgements}
This work was supported by the Federal Ministry of Education and Research (BMBF) for the Berlin Big Data
Center BBDC (01IS14013A). Additional support was provided by the European
Union’s Horizon 2020 research and innovation program under the Marie Sk\l{}odowska-Curie grant agreement NO 792572, the BK21 program funded by Korean National Research Foundation grant (No. 2012-005741).
This research was also supported by Institute for Information \& Communications Technology Promotion and funded by the Korea government (MSIT) (No. 2017-0-00451, No. 2017-0-01779).
A.T. acknowledges support from the European Research Council (ERC-CoG grant BeStMo).
Correspondence to KTS, AT and KRM.

\bibliography{literature}

\providecommand{\noopsort}[1]{}\providecommand{\singleletter}[1]{#1}%
\begin{thebibliography}{44}%
\makeatletter
\providecommand \@ifxundefined [1]{%
 \@ifx{#1\undefined}
}%
\providecommand \@ifnum [1]{%
 \ifnum #1\expandafter \@firstoftwo
 \else \expandafter \@secondoftwo
 \fi
}%
\providecommand \@ifx [1]{%
 \ifx #1\expandafter \@firstoftwo
 \else \expandafter \@secondoftwo
 \fi
}%
\providecommand \natexlab [1]{#1}%
\providecommand \enquote  [1]{``#1''}%
\providecommand \bibnamefont  [1]{#1}%
\providecommand \bibfnamefont [1]{#1}%
\providecommand \citenamefont [1]{#1}%
\providecommand \href@noop [0]{\@secondoftwo}%
\providecommand \href [0]{\begingroup \@sanitize@url \@href}%
\providecommand \@href[1]{\@@startlink{#1}\@@href}%
\providecommand \@@href[1]{\endgroup#1\@@endlink}%
\providecommand \@sanitize@url [0]{\catcode `\\12\catcode `\$12\catcode
  `\&12\catcode `\#12\catcode `\^12\catcode `\_12\catcode `\%12\relax}%
\providecommand \@@startlink[1]{}%
\providecommand \@@endlink[0]{}%
\providecommand \url  [0]{\begingroup\@sanitize@url \@url }%
\providecommand \@url [1]{\endgroup\@href {#1}{\urlprefix }}%
\providecommand \urlprefix  [0]{URL }%
\providecommand \Eprint [0]{\href }%
\providecommand \doibase [0]{http://dx.doi.org/}%
\providecommand \selectlanguage [0]{\@gobble}%
\providecommand \bibinfo  [0]{\@secondoftwo}%
\providecommand \bibfield  [0]{\@secondoftwo}%
\providecommand \translation [1]{[#1]}%
\providecommand \BibitemOpen [0]{}%
\providecommand \bibitemStop [0]{}%
\providecommand \bibitemNoStop [0]{.\EOS\space}%
\providecommand \EOS [0]{\spacefactor3000\relax}%
\providecommand \BibitemShut  [1]{\csname bibitem#1\endcsname}%
\let\auto@bib@innerbib\@empty
\bibitem [{\citenamefont {Bart{\'o}k}\ \emph {et~al.}(2010)\citenamefont
  {Bart{\'o}k}, \citenamefont {Payne}, \citenamefont {Kondor},\ and\
  \citenamefont {Cs{\'a}nyi}}]{bartok2010gaussian}%
  \BibitemOpen
  \bibfield  {author} {\bibinfo {author} {\bibfnamefont {A.~P.}\ \bibnamefont
  {Bart{\'o}k}}, \bibinfo {author} {\bibfnamefont {M.~C.}\ \bibnamefont
  {Payne}}, \bibinfo {author} {\bibfnamefont {R.}~\bibnamefont {Kondor}}, \
  and\ \bibinfo {author} {\bibfnamefont {G.}~\bibnamefont {Cs{\'a}nyi}},\
  }\href@noop {} {\bibfield  {journal} {\bibinfo  {journal} {Physical review
  letters}\ }\textbf {\bibinfo {volume} {104}},\ \bibinfo {pages} {136403}
  (\bibinfo {year} {2010})}\BibitemShut {NoStop}%
\bibitem [{\citenamefont {Rupp}\ \emph {et~al.}(2012)\citenamefont {Rupp},
  \citenamefont {Tkatchenko}, \citenamefont {M{\"u}ller},\ and\ \citenamefont
  {Von~Lilienfeld}}]{rupp2012fast}%
  \BibitemOpen
  \bibfield  {author} {\bibinfo {author} {\bibfnamefont {M.}~\bibnamefont
  {Rupp}}, \bibinfo {author} {\bibfnamefont {A.}~\bibnamefont {Tkatchenko}},
  \bibinfo {author} {\bibfnamefont {K.-R.}\ \bibnamefont {M{\"u}ller}}, \ and\
  \bibinfo {author} {\bibfnamefont {O.~A.}\ \bibnamefont {Von~Lilienfeld}},\
  }\href@noop {} {\bibfield  {journal} {\bibinfo  {journal} {Phys. Rev. Lett.}\
  }\textbf {\bibinfo {volume} {108}},\ \bibinfo {pages} {058301} (\bibinfo
  {year} {2012})}\BibitemShut {NoStop}%
\bibitem [{\citenamefont {Sch{\"u}tt}\ \emph {et~al.}(2014)\citenamefont
  {Sch{\"u}tt}, \citenamefont {Glawe}, \citenamefont {Brockherde},
  \citenamefont {Sanna}, \citenamefont {M{\"u}ller},\ and\ \citenamefont
  {Gross}}]{schutt2014represent}%
  \BibitemOpen
  \bibfield  {author} {\bibinfo {author} {\bibfnamefont {K.~T.}\ \bibnamefont
  {Sch{\"u}tt}}, \bibinfo {author} {\bibfnamefont {H.}~\bibnamefont {Glawe}},
  \bibinfo {author} {\bibfnamefont {F.}~\bibnamefont {Brockherde}}, \bibinfo
  {author} {\bibfnamefont {A.}~\bibnamefont {Sanna}}, \bibinfo {author}
  {\bibfnamefont {K.-R.}\ \bibnamefont {M{\"u}ller}}, \ and\ \bibinfo {author}
  {\bibfnamefont {E.}~\bibnamefont {Gross}},\ }\href@noop {} {\bibfield
  {journal} {\bibinfo  {journal} {Physical Review B}\ }\textbf {\bibinfo
  {volume} {89}},\ \bibinfo {pages} {205118} (\bibinfo {year}
  {2014})}\BibitemShut {NoStop}%
\bibitem [{\citenamefont {Behler}(2015)}]{behler2015constructing}%
  \BibitemOpen
  \bibfield  {author} {\bibinfo {author} {\bibfnamefont {J.}~\bibnamefont
  {Behler}},\ }\href@noop {} {\bibfield  {journal} {\bibinfo  {journal}
  {International Journal of Quantum Chemistry}\ }\textbf {\bibinfo {volume}
  {115}},\ \bibinfo {pages} {1032} (\bibinfo {year} {2015})}\BibitemShut
  {NoStop}%
\bibitem [{\citenamefont {Huo}\ and\ \citenamefont
  {Rupp}(2017)}]{huo2017unified}%
  \BibitemOpen
  \bibfield  {author} {\bibinfo {author} {\bibfnamefont {H.}~\bibnamefont
  {Huo}}\ and\ \bibinfo {author} {\bibfnamefont {M.}~\bibnamefont {Rupp}},\
  }\href@noop {} {\bibfield  {journal} {\bibinfo  {journal} {arXiv preprint
  arXiv:1704.06439}\ } (\bibinfo {year} {2017})}\BibitemShut {NoStop}%
\bibitem [{\citenamefont {Faber}\ \emph {et~al.}(2018)\citenamefont {Faber},
  \citenamefont {Christensen}, \citenamefont {Huang},\ and\ \citenamefont {von
  Lilienfeld}}]{faber2018alchemical}%
  \BibitemOpen
  \bibfield  {author} {\bibinfo {author} {\bibfnamefont {F.~A.}\ \bibnamefont
  {Faber}}, \bibinfo {author} {\bibfnamefont {A.~S.}\ \bibnamefont
  {Christensen}}, \bibinfo {author} {\bibfnamefont {B.}~\bibnamefont {Huang}},
  \ and\ \bibinfo {author} {\bibfnamefont {O.~A.}\ \bibnamefont {von
  Lilienfeld}},\ }\href@noop {} {\bibfield  {journal} {\bibinfo  {journal} {The
  Journal of Chemical Physics}\ }\textbf {\bibinfo {volume} {148}},\ \bibinfo
  {pages} {241717} (\bibinfo {year} {2018})}\BibitemShut {NoStop}%
\bibitem [{\citenamefont {De}\ \emph {et~al.}(2016)\citenamefont {De},
  \citenamefont {Bart{\'o}k}, \citenamefont {Cs{\'a}nyi},\ and\ \citenamefont
  {Ceriotti}}]{de2016comparing}%
  \BibitemOpen
  \bibfield  {author} {\bibinfo {author} {\bibfnamefont {S.}~\bibnamefont
  {De}}, \bibinfo {author} {\bibfnamefont {A.~P.}\ \bibnamefont {Bart{\'o}k}},
  \bibinfo {author} {\bibfnamefont {G.}~\bibnamefont {Cs{\'a}nyi}}, \ and\
  \bibinfo {author} {\bibfnamefont {M.}~\bibnamefont {Ceriotti}},\ }\href@noop
  {} {\bibfield  {journal} {\bibinfo  {journal} {Physical Chemistry Chemical
  Physics}\ }\textbf {\bibinfo {volume} {18}},\ \bibinfo {pages} {13754}
  (\bibinfo {year} {2016})}\BibitemShut {NoStop}%
\bibitem [{\citenamefont {Morawietz}\ \emph {et~al.}(2016)\citenamefont
  {Morawietz}, \citenamefont {Singraber}, \citenamefont {Dellago},\ and\
  \citenamefont {Behler}}]{morawietz2016van}%
  \BibitemOpen
  \bibfield  {author} {\bibinfo {author} {\bibfnamefont {T.}~\bibnamefont
  {Morawietz}}, \bibinfo {author} {\bibfnamefont {A.}~\bibnamefont
  {Singraber}}, \bibinfo {author} {\bibfnamefont {C.}~\bibnamefont {Dellago}},
  \ and\ \bibinfo {author} {\bibfnamefont {J.}~\bibnamefont {Behler}},\
  }\href@noop {} {\bibfield  {journal} {\bibinfo  {journal} {Proceedings of the
  National Academy of Sciences}\ }\textbf {\bibinfo {volume} {113}},\ \bibinfo
  {pages} {8368} (\bibinfo {year} {2016})}\BibitemShut {NoStop}%
\bibitem [{\citenamefont {Gastegger}\ \emph {et~al.}(2017)\citenamefont
  {Gastegger}, \citenamefont {Behler},\ and\ \citenamefont
  {Marquetand}}]{gastegger2017machine}%
  \BibitemOpen
  \bibfield  {author} {\bibinfo {author} {\bibfnamefont {M.}~\bibnamefont
  {Gastegger}}, \bibinfo {author} {\bibfnamefont {J.}~\bibnamefont {Behler}}, \
  and\ \bibinfo {author} {\bibfnamefont {P.}~\bibnamefont {Marquetand}},\
  }\href@noop {} {\bibfield  {journal} {\bibinfo  {journal} {Chemical science}\
  }\textbf {\bibinfo {volume} {8}},\ \bibinfo {pages} {6924} (\bibinfo {year}
  {2017})}\BibitemShut {NoStop}%
\bibitem [{\citenamefont {Chmiela}\ \emph {et~al.}(2017)\citenamefont
  {Chmiela}, \citenamefont {Tkatchenko}, \citenamefont {Sauceda}, \citenamefont
  {Poltavsky}, \citenamefont {Sch{\"u}tt},\ and\ \citenamefont
  {M{\"u}ller}}]{chmiela2017machine}%
  \BibitemOpen
  \bibfield  {author} {\bibinfo {author} {\bibfnamefont {S.}~\bibnamefont
  {Chmiela}}, \bibinfo {author} {\bibfnamefont {A.}~\bibnamefont {Tkatchenko}},
  \bibinfo {author} {\bibfnamefont {H.~E.}\ \bibnamefont {Sauceda}}, \bibinfo
  {author} {\bibfnamefont {I.}~\bibnamefont {Poltavsky}}, \bibinfo {author}
  {\bibfnamefont {K.~T.}\ \bibnamefont {Sch{\"u}tt}}, \ and\ \bibinfo {author}
  {\bibfnamefont {K.-R.}\ \bibnamefont {M{\"u}ller}},\ }\href@noop {}
  {\bibfield  {journal} {\bibinfo  {journal} {Science advances}\ }\textbf
  {\bibinfo {volume} {3}},\ \bibinfo {pages} {e1603015} (\bibinfo {year}
  {2017})}\BibitemShut {NoStop}%
\bibitem [{\citenamefont {Faber}\ \emph {et~al.}(2017)\citenamefont {Faber},
  \citenamefont {Hutchison}, \citenamefont {Huang}, \citenamefont {Gilmer},
  \citenamefont {Schoenholz}, \citenamefont {Dahl}, \citenamefont {Vinyals},
  \citenamefont {Kearnes}, \citenamefont {Riley},\ and\ \citenamefont {von
  Lilienfeld}}]{faber2017prediction}%
  \BibitemOpen
  \bibfield  {author} {\bibinfo {author} {\bibfnamefont {F.~A.}\ \bibnamefont
  {Faber}}, \bibinfo {author} {\bibfnamefont {L.}~\bibnamefont {Hutchison}},
  \bibinfo {author} {\bibfnamefont {B.}~\bibnamefont {Huang}}, \bibinfo
  {author} {\bibfnamefont {J.}~\bibnamefont {Gilmer}}, \bibinfo {author}
  {\bibfnamefont {S.~S.}\ \bibnamefont {Schoenholz}}, \bibinfo {author}
  {\bibfnamefont {G.~E.}\ \bibnamefont {Dahl}}, \bibinfo {author}
  {\bibfnamefont {O.}~\bibnamefont {Vinyals}}, \bibinfo {author} {\bibfnamefont
  {S.}~\bibnamefont {Kearnes}}, \bibinfo {author} {\bibfnamefont {P.~F.}\
  \bibnamefont {Riley}}, \ and\ \bibinfo {author} {\bibfnamefont {O.~A.}\
  \bibnamefont {von Lilienfeld}},\ }\href@noop {} {\bibfield  {journal}
  {\bibinfo  {journal} {Journal of chemical theory and computation}\ }\textbf
  {\bibinfo {volume} {13}},\ \bibinfo {pages} {5255} (\bibinfo {year}
  {2017})}\BibitemShut {NoStop}%
\bibitem [{\citenamefont {Podryabinkin}\ and\ \citenamefont
  {Shapeev}(2017)}]{podryabinkin2017active}%
  \BibitemOpen
  \bibfield  {author} {\bibinfo {author} {\bibfnamefont {E.~V.}\ \bibnamefont
  {Podryabinkin}}\ and\ \bibinfo {author} {\bibfnamefont {A.~V.}\ \bibnamefont
  {Shapeev}},\ }\href@noop {} {\bibfield  {journal} {\bibinfo  {journal}
  {Computational Materials Science}\ }\textbf {\bibinfo {volume} {140}},\
  \bibinfo {pages} {171} (\bibinfo {year} {2017})}\BibitemShut {NoStop}%
\bibitem [{\citenamefont {Brockherde}\ \emph {et~al.}(2017)\citenamefont
  {Brockherde}, \citenamefont {Vogt}, \citenamefont {Li}, \citenamefont
  {Tuckerman}, \citenamefont {Burke},\ and\ \citenamefont
  {M{\"u}ller}}]{brockherde2017bypassing}%
  \BibitemOpen
  \bibfield  {author} {\bibinfo {author} {\bibfnamefont {F.}~\bibnamefont
  {Brockherde}}, \bibinfo {author} {\bibfnamefont {L.}~\bibnamefont {Vogt}},
  \bibinfo {author} {\bibfnamefont {L.}~\bibnamefont {Li}}, \bibinfo {author}
  {\bibfnamefont {M.~E.}\ \bibnamefont {Tuckerman}}, \bibinfo {author}
  {\bibfnamefont {K.}~\bibnamefont {Burke}}, \ and\ \bibinfo {author}
  {\bibfnamefont {K.-R.}\ \bibnamefont {M{\"u}ller}},\ }\href@noop {}
  {\bibfield  {journal} {\bibinfo  {journal} {Nature communications}\ }\textbf
  {\bibinfo {volume} {8}},\ \bibinfo {pages} {872} (\bibinfo {year}
  {2017})}\BibitemShut {NoStop}%
\bibitem [{\citenamefont {Bart{\'o}k}\ \emph {et~al.}(2017)\citenamefont
  {Bart{\'o}k}, \citenamefont {De}, \citenamefont {Poelking}, \citenamefont
  {Bernstein}, \citenamefont {Kermode}, \citenamefont {Cs{\'a}nyi},\ and\
  \citenamefont {Ceriotti}}]{bartok2017machine}%
  \BibitemOpen
  \bibfield  {author} {\bibinfo {author} {\bibfnamefont {A.~P.}\ \bibnamefont
  {Bart{\'o}k}}, \bibinfo {author} {\bibfnamefont {S.}~\bibnamefont {De}},
  \bibinfo {author} {\bibfnamefont {C.}~\bibnamefont {Poelking}}, \bibinfo
  {author} {\bibfnamefont {N.}~\bibnamefont {Bernstein}}, \bibinfo {author}
  {\bibfnamefont {J.~R.}\ \bibnamefont {Kermode}}, \bibinfo {author}
  {\bibfnamefont {G.}~\bibnamefont {Cs{\'a}nyi}}, \ and\ \bibinfo {author}
  {\bibfnamefont {M.}~\bibnamefont {Ceriotti}},\ }\href@noop {} {\bibfield
  {journal} {\bibinfo  {journal} {Science advances}\ }\textbf {\bibinfo
  {volume} {3}},\ \bibinfo {pages} {e1701816} (\bibinfo {year}
  {2017})}\BibitemShut {NoStop}%
\bibitem [{\citenamefont {Sch{\"u}tt}\ \emph {et~al.}(2018)\citenamefont
  {Sch{\"u}tt}, \citenamefont {Sauceda}, \citenamefont {Kindermans},
  \citenamefont {Tkatchenko},\ and\ \citenamefont
  {M{\"u}ller}}]{schutt2018schnet}%
  \BibitemOpen
  \bibfield  {author} {\bibinfo {author} {\bibfnamefont {K.~T.}\ \bibnamefont
  {Sch{\"u}tt}}, \bibinfo {author} {\bibfnamefont {H.~E.}\ \bibnamefont
  {Sauceda}}, \bibinfo {author} {\bibfnamefont {P.-J.}\ \bibnamefont
  {Kindermans}}, \bibinfo {author} {\bibfnamefont {A.}~\bibnamefont
  {Tkatchenko}}, \ and\ \bibinfo {author} {\bibfnamefont {K.-R.}\ \bibnamefont
  {M{\"u}ller}},\ }\href@noop {} {\bibfield  {journal} {\bibinfo  {journal}
  {The Journal of Chemical Physics}\ }\textbf {\bibinfo {volume} {148}},\
  \bibinfo {pages} {241722} (\bibinfo {year} {2018})}\BibitemShut {NoStop}%
\bibitem [{\citenamefont {Chmiela}\ \emph {et~al.}(2018)\citenamefont
  {Chmiela}, \citenamefont {Sauceda}, \citenamefont {M{\"u}ller},\ and\
  \citenamefont {Tkatchenko}}]{chmiela2018towards}%
  \BibitemOpen
  \bibfield  {author} {\bibinfo {author} {\bibfnamefont {S.}~\bibnamefont
  {Chmiela}}, \bibinfo {author} {\bibfnamefont {H.~E.}\ \bibnamefont
  {Sauceda}}, \bibinfo {author} {\bibfnamefont {K.-R.}\ \bibnamefont
  {M{\"u}ller}}, \ and\ \bibinfo {author} {\bibfnamefont {A.}~\bibnamefont
  {Tkatchenko}},\ }\href@noop {} {\bibfield  {journal} {\bibinfo  {journal}
  {arXiv preprint arXiv:1802.09238}\ } (\bibinfo {year} {2018})}\BibitemShut
  {NoStop}%
\bibitem [{\citenamefont {Ziletti}\ \emph {et~al.}(2018)\citenamefont
  {Ziletti}, \citenamefont {Kumar}, \citenamefont {Scheffler},\ and\
  \citenamefont {Ghiringhelli}}]{ziletti2018insightful}%
  \BibitemOpen
  \bibfield  {author} {\bibinfo {author} {\bibfnamefont {A.}~\bibnamefont
  {Ziletti}}, \bibinfo {author} {\bibfnamefont {D.}~\bibnamefont {Kumar}},
  \bibinfo {author} {\bibfnamefont {M.}~\bibnamefont {Scheffler}}, \ and\
  \bibinfo {author} {\bibfnamefont {L.~M.}\ \bibnamefont {Ghiringhelli}},\
  }\href@noop {} {\bibfield  {journal} {\bibinfo  {journal} {Nature
  communications}\ }\textbf {\bibinfo {volume} {9}},\ \bibinfo {pages} {2775}
  (\bibinfo {year} {2018})}\BibitemShut {NoStop}%
\bibitem [{\citenamefont {Dragoni}\ \emph {et~al.}(2018)\citenamefont
  {Dragoni}, \citenamefont {Daff}, \citenamefont {Cs{\'a}nyi},\ and\
  \citenamefont {Marzari}}]{dragoni2018achieving}%
  \BibitemOpen
  \bibfield  {author} {\bibinfo {author} {\bibfnamefont {D.}~\bibnamefont
  {Dragoni}}, \bibinfo {author} {\bibfnamefont {T.~D.}\ \bibnamefont {Daff}},
  \bibinfo {author} {\bibfnamefont {G.}~\bibnamefont {Cs{\'a}nyi}}, \ and\
  \bibinfo {author} {\bibfnamefont {N.}~\bibnamefont {Marzari}},\ }\href@noop
  {} {\bibfield  {journal} {\bibinfo  {journal} {Physical Review Materials}\
  }\textbf {\bibinfo {volume} {2}},\ \bibinfo {pages} {013808} (\bibinfo {year}
  {2018})}\BibitemShut {NoStop}%
\bibitem [{\citenamefont {Behler}\ and\ \citenamefont
  {Parrinello}(2007)}]{behler2007generalized}%
  \BibitemOpen
  \bibfield  {author} {\bibinfo {author} {\bibfnamefont {J.}~\bibnamefont
  {Behler}}\ and\ \bibinfo {author} {\bibfnamefont {M.}~\bibnamefont
  {Parrinello}},\ }\href@noop {} {\bibfield  {journal} {\bibinfo  {journal}
  {Physical Review Letters}\ }\textbf {\bibinfo {volume} {98}},\ \bibinfo
  {pages} {146401} (\bibinfo {year} {2007})}\BibitemShut {NoStop}%
\bibitem [{\citenamefont {Montavon}\ \emph {et~al.}(2012)\citenamefont
  {Montavon}, \citenamefont {Hansen}, \citenamefont {Fazli}, \citenamefont
  {Rupp}, \citenamefont {Biegler}, \citenamefont {Ziehe}, \citenamefont
  {Tkatchenko}, \citenamefont {Lilienfeld},\ and\ \citenamefont
  {M{\"u}ller}}]{montavon2012learning}%
  \BibitemOpen
  \bibfield  {author} {\bibinfo {author} {\bibfnamefont {G.}~\bibnamefont
  {Montavon}}, \bibinfo {author} {\bibfnamefont {K.}~\bibnamefont {Hansen}},
  \bibinfo {author} {\bibfnamefont {S.}~\bibnamefont {Fazli}}, \bibinfo
  {author} {\bibfnamefont {M.}~\bibnamefont {Rupp}}, \bibinfo {author}
  {\bibfnamefont {F.}~\bibnamefont {Biegler}}, \bibinfo {author} {\bibfnamefont
  {A.}~\bibnamefont {Ziehe}}, \bibinfo {author} {\bibfnamefont
  {A.}~\bibnamefont {Tkatchenko}}, \bibinfo {author} {\bibfnamefont {A.~V.}\
  \bibnamefont {Lilienfeld}}, \ and\ \bibinfo {author} {\bibfnamefont {K.-R.}\
  \bibnamefont {M{\"u}ller}},\ }in\ \href@noop {} {\emph {\bibinfo {booktitle}
  {Advances in Neural Information Processing Systems}}}\ (\bibinfo {year}
  {2012})\ pp.\ \bibinfo {pages} {440--448}\BibitemShut {NoStop}%
\bibitem [{\citenamefont {Montavon}\ \emph {et~al.}(2013)\citenamefont
  {Montavon}, \citenamefont {Rupp}, \citenamefont {Gobre}, \citenamefont
  {Vazquez-Mayagoitia}, \citenamefont {Hansen}, \citenamefont {Tkatchenko},
  \citenamefont {M{\"u}ller},\ and\ \citenamefont
  {Von~Lilienfeld}}]{montavon2013machine}%
  \BibitemOpen
  \bibfield  {author} {\bibinfo {author} {\bibfnamefont {G.}~\bibnamefont
  {Montavon}}, \bibinfo {author} {\bibfnamefont {M.}~\bibnamefont {Rupp}},
  \bibinfo {author} {\bibfnamefont {V.}~\bibnamefont {Gobre}}, \bibinfo
  {author} {\bibfnamefont {A.}~\bibnamefont {Vazquez-Mayagoitia}}, \bibinfo
  {author} {\bibfnamefont {K.}~\bibnamefont {Hansen}}, \bibinfo {author}
  {\bibfnamefont {A.}~\bibnamefont {Tkatchenko}}, \bibinfo {author}
  {\bibfnamefont {K.-R.}\ \bibnamefont {M{\"u}ller}}, \ and\ \bibinfo {author}
  {\bibfnamefont {O.~A.}\ \bibnamefont {Von~Lilienfeld}},\ }\href@noop {}
  {\bibfield  {journal} {\bibinfo  {journal} {New Journal of Physics}\ }\textbf
  {\bibinfo {volume} {15}},\ \bibinfo {pages} {095003} (\bibinfo {year}
  {2013})}\BibitemShut {NoStop}%
\bibitem [{\citenamefont {Zhang}\ \emph {et~al.}(2018)\citenamefont {Zhang},
  \citenamefont {Han}, \citenamefont {Wang}, \citenamefont {Car},\ and\
  \citenamefont {Weinan}}]{zhang2018deep}%
  \BibitemOpen
  \bibfield  {author} {\bibinfo {author} {\bibfnamefont {L.}~\bibnamefont
  {Zhang}}, \bibinfo {author} {\bibfnamefont {J.}~\bibnamefont {Han}}, \bibinfo
  {author} {\bibfnamefont {H.}~\bibnamefont {Wang}}, \bibinfo {author}
  {\bibfnamefont {R.}~\bibnamefont {Car}}, \ and\ \bibinfo {author}
  {\bibfnamefont {E.}~\bibnamefont {Weinan}},\ }\href@noop {} {\bibfield
  {journal} {\bibinfo  {journal} {Physical review letters}\ }\textbf {\bibinfo
  {volume} {120}},\ \bibinfo {pages} {143001} (\bibinfo {year}
  {2018})}\BibitemShut {NoStop}%
\bibitem [{\citenamefont {Smith}\ \emph
  {et~al.}(2017{\natexlab{a}})\citenamefont {Smith}, \citenamefont {Isayev},\
  and\ \citenamefont {Roitberg}}]{smith2017ani}%
  \BibitemOpen
  \bibfield  {author} {\bibinfo {author} {\bibfnamefont {J.~S.}\ \bibnamefont
  {Smith}}, \bibinfo {author} {\bibfnamefont {O.}~\bibnamefont {Isayev}}, \
  and\ \bibinfo {author} {\bibfnamefont {A.~E.}\ \bibnamefont {Roitberg}},\
  }\href@noop {} {\bibfield  {journal} {\bibinfo  {journal} {Chemical science}\
  }\textbf {\bibinfo {volume} {8}},\ \bibinfo {pages} {3192} (\bibinfo {year}
  {2017}{\natexlab{a}})}\BibitemShut {NoStop}%
\bibitem [{\citenamefont {Gastegger}\ \emph {et~al.}(2018)\citenamefont
  {Gastegger}, \citenamefont {Schwiedrzik}, \citenamefont {Bittermann},
  \citenamefont {Berzsenyi},\ and\ \citenamefont
  {Marquetand}}]{gastegger2018wacsf}%
  \BibitemOpen
  \bibfield  {author} {\bibinfo {author} {\bibfnamefont {M.}~\bibnamefont
  {Gastegger}}, \bibinfo {author} {\bibfnamefont {L.}~\bibnamefont
  {Schwiedrzik}}, \bibinfo {author} {\bibfnamefont {M.}~\bibnamefont
  {Bittermann}}, \bibinfo {author} {\bibfnamefont {F.}~\bibnamefont
  {Berzsenyi}}, \ and\ \bibinfo {author} {\bibfnamefont {P.}~\bibnamefont
  {Marquetand}},\ }\href@noop {} {\bibfield  {journal} {\bibinfo  {journal}
  {The Journal of Chemical Physics}\ }\textbf {\bibinfo {volume} {148}},\
  \bibinfo {pages} {241709} (\bibinfo {year} {2018})}\BibitemShut {NoStop}%
\bibitem [{\citenamefont {Sch{\"u}tt}\ \emph
  {et~al.}(2017{\natexlab{a}})\citenamefont {Sch{\"u}tt}, \citenamefont
  {Arbabzadah}, \citenamefont {Chmiela}, \citenamefont {M{\"u}ller},\ and\
  \citenamefont {Tkatchenko}}]{schuett2017quantum}%
  \BibitemOpen
  \bibfield  {author} {\bibinfo {author} {\bibfnamefont {K.~T.}\ \bibnamefont
  {Sch{\"u}tt}}, \bibinfo {author} {\bibfnamefont {F.}~\bibnamefont
  {Arbabzadah}}, \bibinfo {author} {\bibfnamefont {S.}~\bibnamefont {Chmiela}},
  \bibinfo {author} {\bibfnamefont {K.~R.}\ \bibnamefont {M{\"u}ller}}, \ and\
  \bibinfo {author} {\bibfnamefont {A.}~\bibnamefont {Tkatchenko}},\
  }\href@noop {} {\bibfield  {journal} {\bibinfo  {journal} {Nature
  communications}\ }\textbf {\bibinfo {volume} {8}},\ \bibinfo {pages} {13890}
  (\bibinfo {year} {2017}{\natexlab{a}})}\BibitemShut {NoStop}%
\bibitem [{\citenamefont {Gilmer}\ \emph {et~al.}(2017)\citenamefont {Gilmer},
  \citenamefont {Schoenholz}, \citenamefont {Riley}, \citenamefont {Vinyals},\
  and\ \citenamefont {Dahl}}]{gilmer2017neural}%
  \BibitemOpen
  \bibfield  {author} {\bibinfo {author} {\bibfnamefont {J.}~\bibnamefont
  {Gilmer}}, \bibinfo {author} {\bibfnamefont {S.~S.}\ \bibnamefont
  {Schoenholz}}, \bibinfo {author} {\bibfnamefont {P.~F.}\ \bibnamefont
  {Riley}}, \bibinfo {author} {\bibfnamefont {O.}~\bibnamefont {Vinyals}}, \
  and\ \bibinfo {author} {\bibfnamefont {G.~E.}\ \bibnamefont {Dahl}},\
  }\href@noop {} {\bibfield  {journal} {\bibinfo  {journal} {arXiv preprint
  arXiv:1704.01212}\ } (\bibinfo {year} {2017})}\BibitemShut {NoStop}%
\bibitem [{\citenamefont {Sch{\"u}tt}\ \emph
  {et~al.}(2017{\natexlab{b}})\citenamefont {Sch{\"u}tt}, \citenamefont
  {Kindermans}, \citenamefont {Sauceda}, \citenamefont {Chmiela}, \citenamefont
  {Tkatchenko},\ and\ \citenamefont {M{\"u}ller}}]{schutt2017schnet}%
  \BibitemOpen
  \bibfield  {author} {\bibinfo {author} {\bibfnamefont {K.~T.}\ \bibnamefont
  {Sch{\"u}tt}}, \bibinfo {author} {\bibfnamefont {P.-J.}\ \bibnamefont
  {Kindermans}}, \bibinfo {author} {\bibfnamefont {H.~E.}\ \bibnamefont
  {Sauceda}}, \bibinfo {author} {\bibfnamefont {S.}~\bibnamefont {Chmiela}},
  \bibinfo {author} {\bibfnamefont {A.}~\bibnamefont {Tkatchenko}}, \ and\
  \bibinfo {author} {\bibfnamefont {K.-R.}\ \bibnamefont {M{\"u}ller}},\ }in\
  \href@noop {} {\emph {\bibinfo {booktitle} {Advances in Neural Information
  Processing Systems}}}\ (\bibinfo {year} {2017})\ pp.\ \bibinfo {pages}
  {991--1001}\BibitemShut {NoStop}%
\bibitem [{\citenamefont {Lubbers}\ \emph {et~al.}(2018)\citenamefont
  {Lubbers}, \citenamefont {Smith},\ and\ \citenamefont
  {Barros}}]{lubbers2018hierarchical}%
  \BibitemOpen
  \bibfield  {author} {\bibinfo {author} {\bibfnamefont {N.}~\bibnamefont
  {Lubbers}}, \bibinfo {author} {\bibfnamefont {J.~S.}\ \bibnamefont {Smith}},
  \ and\ \bibinfo {author} {\bibfnamefont {K.}~\bibnamefont {Barros}},\
  }\href@noop {} {\bibfield  {journal} {\bibinfo  {journal} {The Journal of
  Chemical Physics}\ }\textbf {\bibinfo {volume} {148}},\ \bibinfo {pages}
  {241715} (\bibinfo {year} {2018})}\BibitemShut {NoStop}%
\bibitem [{\citenamefont {Behler}(2011)}]{behler2011atom}%
  \BibitemOpen
  \bibfield  {author} {\bibinfo {author} {\bibfnamefont {J.}~\bibnamefont
  {Behler}},\ }\href@noop {} {\bibfield  {journal} {\bibinfo  {journal} {The
  Journal of Chemical Physics}\ }\textbf {\bibinfo {volume} {134}},\ \bibinfo
  {pages} {074106} (\bibinfo {year} {2011})}\BibitemShut {NoStop}%
\bibitem [{\citenamefont {Behler}(2017)}]{behler2017first}%
  \BibitemOpen
  \bibfield  {author} {\bibinfo {author} {\bibfnamefont {J.}~\bibnamefont
  {Behler}},\ }\href@noop {} {\bibfield  {journal} {\bibinfo  {journal}
  {Angewandte Chemie International Edition}\ }\textbf {\bibinfo {volume}
  {56}},\ \bibinfo {pages} {12828} (\bibinfo {year} {2017})}\BibitemShut
  {NoStop}%
\bibitem [{\citenamefont {He}\ \emph {et~al.}(2016)\citenamefont {He},
  \citenamefont {Zhang}, \citenamefont {Ren},\ and\ \citenamefont
  {Sun}}]{he2016deep}%
  \BibitemOpen
  \bibfield  {author} {\bibinfo {author} {\bibfnamefont {K.}~\bibnamefont
  {He}}, \bibinfo {author} {\bibfnamefont {X.}~\bibnamefont {Zhang}}, \bibinfo
  {author} {\bibfnamefont {S.}~\bibnamefont {Ren}}, \ and\ \bibinfo {author}
  {\bibfnamefont {J.}~\bibnamefont {Sun}},\ }in\ \href@noop {} {\emph {\bibinfo
  {booktitle} {Proceedings of the IEEE conference on computer vision and
  pattern recognition}}}\ (\bibinfo {year} {2016})\ pp.\ \bibinfo {pages}
  {770--778}\BibitemShut {NoStop}%
\bibitem [{\citenamefont {Larsen}\ \emph {et~al.}(2017)\citenamefont {Larsen},
  \citenamefont {Mortensen}, \citenamefont {Blomqvist}, \citenamefont
  {Castelli}, \citenamefont {Christensen}, \citenamefont {Dułak},
  \citenamefont {Friis}, \citenamefont {Groves}, \citenamefont {Hammer},
  \citenamefont {Hargus}, \citenamefont {Hermes}, \citenamefont {Jennings},
  \citenamefont {Jensen}, \citenamefont {Kermode}, \citenamefont {Kitchin},
  \citenamefont {Kolsbjerg}, \citenamefont {Kubal}, \citenamefont {Kaasbjerg},
  \citenamefont {Lysgaard}, \citenamefont {Maronsson}, \citenamefont {Maxson},
  \citenamefont {Olsen}, \citenamefont {Pastewka}, \citenamefont {Peterson},
  \citenamefont {Rostgaard}, \citenamefont {Schiøtz}, \citenamefont {Schütt},
  \citenamefont {Strange}, \citenamefont {Thygesen}, \citenamefont {Vegge},
  \citenamefont {Vilhelmsen}, \citenamefont {Walter}, \citenamefont {Zeng},\
  and\ \citenamefont {Jacobsen}}]{ase}%
  \BibitemOpen
  \bibfield  {author} {\bibinfo {author} {\bibfnamefont {A.~H.}\ \bibnamefont
  {Larsen}}, \bibinfo {author} {\bibfnamefont {J.~J.}\ \bibnamefont
  {Mortensen}}, \bibinfo {author} {\bibfnamefont {J.}~\bibnamefont
  {Blomqvist}}, \bibinfo {author} {\bibfnamefont {I.~E.}\ \bibnamefont
  {Castelli}}, \bibinfo {author} {\bibfnamefont {R.}~\bibnamefont
  {Christensen}}, \bibinfo {author} {\bibfnamefont {M.}~\bibnamefont {Dułak}},
  \bibinfo {author} {\bibfnamefont {J.}~\bibnamefont {Friis}}, \bibinfo
  {author} {\bibfnamefont {M.~N.}\ \bibnamefont {Groves}}, \bibinfo {author}
  {\bibfnamefont {B.}~\bibnamefont {Hammer}}, \bibinfo {author} {\bibfnamefont
  {C.}~\bibnamefont {Hargus}}, \bibinfo {author} {\bibfnamefont {E.~D.}\
  \bibnamefont {Hermes}}, \bibinfo {author} {\bibfnamefont {P.~C.}\
  \bibnamefont {Jennings}}, \bibinfo {author} {\bibfnamefont {P.~B.}\
  \bibnamefont {Jensen}}, \bibinfo {author} {\bibfnamefont {J.}~\bibnamefont
  {Kermode}}, \bibinfo {author} {\bibfnamefont {J.~R.}\ \bibnamefont
  {Kitchin}}, \bibinfo {author} {\bibfnamefont {E.~L.}\ \bibnamefont
  {Kolsbjerg}}, \bibinfo {author} {\bibfnamefont {J.}~\bibnamefont {Kubal}},
  \bibinfo {author} {\bibfnamefont {K.}~\bibnamefont {Kaasbjerg}}, \bibinfo
  {author} {\bibfnamefont {S.}~\bibnamefont {Lysgaard}}, \bibinfo {author}
  {\bibfnamefont {J.~B.}\ \bibnamefont {Maronsson}}, \bibinfo {author}
  {\bibfnamefont {T.}~\bibnamefont {Maxson}}, \bibinfo {author} {\bibfnamefont
  {T.}~\bibnamefont {Olsen}}, \bibinfo {author} {\bibfnamefont
  {L.}~\bibnamefont {Pastewka}}, \bibinfo {author} {\bibfnamefont
  {A.}~\bibnamefont {Peterson}}, \bibinfo {author} {\bibfnamefont
  {C.}~\bibnamefont {Rostgaard}}, \bibinfo {author} {\bibfnamefont
  {J.}~\bibnamefont {Schiøtz}}, \bibinfo {author} {\bibfnamefont
  {O.}~\bibnamefont {Schütt}}, \bibinfo {author} {\bibfnamefont
  {M.}~\bibnamefont {Strange}}, \bibinfo {author} {\bibfnamefont {K.~S.}\
  \bibnamefont {Thygesen}}, \bibinfo {author} {\bibfnamefont {T.}~\bibnamefont
  {Vegge}}, \bibinfo {author} {\bibfnamefont {L.}~\bibnamefont {Vilhelmsen}},
  \bibinfo {author} {\bibfnamefont {M.}~\bibnamefont {Walter}}, \bibinfo
  {author} {\bibfnamefont {Z.}~\bibnamefont {Zeng}}, \ and\ \bibinfo {author}
  {\bibfnamefont {K.~W.}\ \bibnamefont {Jacobsen}},\ }\href
  {http://stacks.iop.org/0953-8984/29/i=27/a=273002} {\bibfield  {journal}
  {\bibinfo  {journal} {Journal of Physics: Condensed Matter}\ }\textbf
  {\bibinfo {volume} {29}},\ \bibinfo {pages} {273002} (\bibinfo {year}
  {2017})}\BibitemShut {NoStop}%
\bibitem [{\citenamefont {Ruddigkeit}\ \emph {et~al.}(2012)\citenamefont
  {Ruddigkeit}, \citenamefont {van Deursen}, \citenamefont {Blum},\ and\
  \citenamefont {Reymond}}]{qm9one}%
  \BibitemOpen
  \bibfield  {author} {\bibinfo {author} {\bibfnamefont {L.}~\bibnamefont
  {Ruddigkeit}}, \bibinfo {author} {\bibfnamefont {R.}~\bibnamefont {van
  Deursen}}, \bibinfo {author} {\bibfnamefont {L.~C.}\ \bibnamefont {Blum}}, \
  and\ \bibinfo {author} {\bibfnamefont {J.-L.}\ \bibnamefont {Reymond}},\
  }\href {\doibase 10.1021/ci300415d} {\bibfield  {journal} {\bibinfo
  {journal} {Journal of Chemical Information and Modeling}\ }\textbf {\bibinfo
  {volume} {52}},\ \bibinfo {pages} {2864} (\bibinfo {year} {2012})},\ \bibinfo
  {note} {pMID: 23088335},\ \Eprint
  {http://arxiv.org/abs/https://doi.org/10.1021/ci300415d}
  {https://doi.org/10.1021/ci300415d} \BibitemShut {NoStop}%
\bibitem [{\citenamefont {Ramakrishnan}\ \emph {et~al.}(2014)\citenamefont
  {Ramakrishnan}, \citenamefont {Dral}, \citenamefont {Rupp},\ and\
  \citenamefont {von Lilienfeld}}]{qm9two}%
  \BibitemOpen
  \bibfield  {author} {\bibinfo {author} {\bibfnamefont {R.}~\bibnamefont
  {Ramakrishnan}}, \bibinfo {author} {\bibfnamefont {P.~O.}\ \bibnamefont
  {Dral}}, \bibinfo {author} {\bibfnamefont {M.}~\bibnamefont {Rupp}}, \ and\
  \bibinfo {author} {\bibfnamefont {O.~A.}\ \bibnamefont {von Lilienfeld}},\
  }\href@noop {} {\bibfield  {journal} {\bibinfo  {journal} {Scientific Data}\
  }\textbf {\bibinfo {volume} {1}} (\bibinfo {year} {2014})}\BibitemShut
  {NoStop}%
\bibitem [{\citenamefont {Smith}\ \emph
  {et~al.}(2017{\natexlab{b}})\citenamefont {Smith}, \citenamefont {Isayev},\
  and\ \citenamefont {Roitberg}}]{ani1}%
  \BibitemOpen
  \bibfield  {author} {\bibinfo {author} {\bibfnamefont {J.~S.}\ \bibnamefont
  {Smith}}, \bibinfo {author} {\bibfnamefont {O.}~\bibnamefont {Isayev}}, \
  and\ \bibinfo {author} {\bibfnamefont {A.~E.}\ \bibnamefont {Roitberg}},\
  }\href@noop {} {\bibfield  {journal} {\bibinfo  {journal} {Scientific Data}\
  }\textbf {\bibinfo {volume} {4}},\ \bibinfo {pages} {170193} (\bibinfo {year}
  {2017}{\natexlab{b}})}\BibitemShut {NoStop}%
\bibitem [{\citenamefont {Jain}\ \emph {et~al.}(2013)\citenamefont {Jain},
  \citenamefont {Ong}, \citenamefont {Hautier}, \citenamefont {Chen},
  \citenamefont {Richards}, \citenamefont {Dacek}, \citenamefont {Cholia},
  \citenamefont {Gunter}, \citenamefont {Skinner}, \citenamefont {Ceder} \emph
  {et~al.}}]{mp1}%
  \BibitemOpen
  \bibfield  {author} {\bibinfo {author} {\bibfnamefont {A.}~\bibnamefont
  {Jain}}, \bibinfo {author} {\bibfnamefont {S.~P.}\ \bibnamefont {Ong}},
  \bibinfo {author} {\bibfnamefont {G.}~\bibnamefont {Hautier}}, \bibinfo
  {author} {\bibfnamefont {W.}~\bibnamefont {Chen}}, \bibinfo {author}
  {\bibfnamefont {W.~D.}\ \bibnamefont {Richards}}, \bibinfo {author}
  {\bibfnamefont {S.}~\bibnamefont {Dacek}}, \bibinfo {author} {\bibfnamefont
  {S.}~\bibnamefont {Cholia}}, \bibinfo {author} {\bibfnamefont
  {D.}~\bibnamefont {Gunter}}, \bibinfo {author} {\bibfnamefont
  {D.}~\bibnamefont {Skinner}}, \bibinfo {author} {\bibfnamefont
  {G.}~\bibnamefont {Ceder}},  \emph {et~al.},\ }\href@noop {} {\bibfield
  {journal} {\bibinfo  {journal} {Apl Materials}\ }\textbf {\bibinfo {volume}
  {1}},\ \bibinfo {pages} {011002} (\bibinfo {year} {2013})}\BibitemShut
  {NoStop}%
\bibitem [{\citenamefont {Abadi}\ \emph {et~al.}(2015)\citenamefont {Abadi},
  \citenamefont {Agarwal}, \citenamefont {Barham}, \citenamefont {Brevdo},
  \citenamefont {Chen}, \citenamefont {Citro}, \citenamefont {Corrado},
  \citenamefont {Davis}, \citenamefont {Dean}, \citenamefont {Devin},
  \citenamefont {Ghemawat}, \citenamefont {Goodfellow}, \citenamefont {Harp},
  \citenamefont {Irving}, \citenamefont {Isard}, \citenamefont {Jia},
  \citenamefont {Jozefowicz}, \citenamefont {Kaiser}, \citenamefont {Kudlur},
  \citenamefont {Levenberg}, \citenamefont {Man\'{e}}, \citenamefont {Monga},
  \citenamefont {Moore}, \citenamefont {Murray}, \citenamefont {Olah},
  \citenamefont {Schuster}, \citenamefont {Shlens}, \citenamefont {Steiner},
  \citenamefont {Sutskever}, \citenamefont {Talwar}, \citenamefont {Tucker},
  \citenamefont {Vanhoucke}, \citenamefont {Vasudevan}, \citenamefont
  {Vi\'{e}gas}, \citenamefont {Vinyals}, \citenamefont {Warden}, \citenamefont
  {Wattenberg}, \citenamefont {Wicke}, \citenamefont {Yu},\ and\ \citenamefont
  {Zheng}}]{tensorflow}%
  \BibitemOpen
  \bibfield  {author} {\bibinfo {author} {\bibfnamefont {M.}~\bibnamefont
  {Abadi}}, \bibinfo {author} {\bibfnamefont {A.}~\bibnamefont {Agarwal}},
  \bibinfo {author} {\bibfnamefont {P.}~\bibnamefont {Barham}}, \bibinfo
  {author} {\bibfnamefont {E.}~\bibnamefont {Brevdo}}, \bibinfo {author}
  {\bibfnamefont {Z.}~\bibnamefont {Chen}}, \bibinfo {author} {\bibfnamefont
  {C.}~\bibnamefont {Citro}}, \bibinfo {author} {\bibfnamefont {G.~S.}\
  \bibnamefont {Corrado}}, \bibinfo {author} {\bibfnamefont {A.}~\bibnamefont
  {Davis}}, \bibinfo {author} {\bibfnamefont {J.}~\bibnamefont {Dean}},
  \bibinfo {author} {\bibfnamefont {M.}~\bibnamefont {Devin}}, \bibinfo
  {author} {\bibfnamefont {S.}~\bibnamefont {Ghemawat}}, \bibinfo {author}
  {\bibfnamefont {I.}~\bibnamefont {Goodfellow}}, \bibinfo {author}
  {\bibfnamefont {A.}~\bibnamefont {Harp}}, \bibinfo {author} {\bibfnamefont
  {G.}~\bibnamefont {Irving}}, \bibinfo {author} {\bibfnamefont
  {M.}~\bibnamefont {Isard}}, \bibinfo {author} {\bibfnamefont
  {Y.}~\bibnamefont {Jia}}, \bibinfo {author} {\bibfnamefont {R.}~\bibnamefont
  {Jozefowicz}}, \bibinfo {author} {\bibfnamefont {L.}~\bibnamefont {Kaiser}},
  \bibinfo {author} {\bibfnamefont {M.}~\bibnamefont {Kudlur}}, \bibinfo
  {author} {\bibfnamefont {J.}~\bibnamefont {Levenberg}}, \bibinfo {author}
  {\bibfnamefont {D.}~\bibnamefont {Man\'{e}}}, \bibinfo {author}
  {\bibfnamefont {R.}~\bibnamefont {Monga}}, \bibinfo {author} {\bibfnamefont
  {S.}~\bibnamefont {Moore}}, \bibinfo {author} {\bibfnamefont
  {D.}~\bibnamefont {Murray}}, \bibinfo {author} {\bibfnamefont
  {C.}~\bibnamefont {Olah}}, \bibinfo {author} {\bibfnamefont {M.}~\bibnamefont
  {Schuster}}, \bibinfo {author} {\bibfnamefont {J.}~\bibnamefont {Shlens}},
  \bibinfo {author} {\bibfnamefont {B.}~\bibnamefont {Steiner}}, \bibinfo
  {author} {\bibfnamefont {I.}~\bibnamefont {Sutskever}}, \bibinfo {author}
  {\bibfnamefont {K.}~\bibnamefont {Talwar}}, \bibinfo {author} {\bibfnamefont
  {P.}~\bibnamefont {Tucker}}, \bibinfo {author} {\bibfnamefont
  {V.}~\bibnamefont {Vanhoucke}}, \bibinfo {author} {\bibfnamefont
  {V.}~\bibnamefont {Vasudevan}}, \bibinfo {author} {\bibfnamefont
  {F.}~\bibnamefont {Vi\'{e}gas}}, \bibinfo {author} {\bibfnamefont
  {O.}~\bibnamefont {Vinyals}}, \bibinfo {author} {\bibfnamefont
  {P.}~\bibnamefont {Warden}}, \bibinfo {author} {\bibfnamefont
  {M.}~\bibnamefont {Wattenberg}}, \bibinfo {author} {\bibfnamefont
  {M.}~\bibnamefont {Wicke}}, \bibinfo {author} {\bibfnamefont
  {Y.}~\bibnamefont {Yu}}, \ and\ \bibinfo {author} {\bibfnamefont
  {X.}~\bibnamefont {Zheng}},\ }\href {https://www.tensorflow.org/} {\enquote
  {\bibinfo {title} {{TensorFlow}: Large-scale machine learning on
  heterogeneous systems},}\ } (\bibinfo {year} {2015}),\ \bibinfo {note}
  {software available from tensorflow.org}\BibitemShut {NoStop}%
\bibitem [{\citenamefont {Paszke}\ \emph {et~al.}(2017)\citenamefont {Paszke},
  \citenamefont {Gross}, \citenamefont {Chintala}, \citenamefont {Chanan},
  \citenamefont {Yang}, \citenamefont {DeVito}, \citenamefont {Lin},
  \citenamefont {Desmaison}, \citenamefont {Antiga},\ and\ \citenamefont
  {Lerer}}]{pytorch}%
  \BibitemOpen
  \bibfield  {author} {\bibinfo {author} {\bibfnamefont {A.}~\bibnamefont
  {Paszke}}, \bibinfo {author} {\bibfnamefont {S.}~\bibnamefont {Gross}},
  \bibinfo {author} {\bibfnamefont {S.}~\bibnamefont {Chintala}}, \bibinfo
  {author} {\bibfnamefont {G.}~\bibnamefont {Chanan}}, \bibinfo {author}
  {\bibfnamefont {E.}~\bibnamefont {Yang}}, \bibinfo {author} {\bibfnamefont
  {Z.}~\bibnamefont {DeVito}}, \bibinfo {author} {\bibfnamefont
  {Z.}~\bibnamefont {Lin}}, \bibinfo {author} {\bibfnamefont {A.}~\bibnamefont
  {Desmaison}}, \bibinfo {author} {\bibfnamefont {L.}~\bibnamefont {Antiga}}, \
  and\ \bibinfo {author} {\bibfnamefont {A.}~\bibnamefont {Lerer}},\
  }\href@noop {} {\  (\bibinfo {year} {2017})}\BibitemShut {NoStop}%
\bibitem [{\citenamefont {Jones}\ \emph {et~al.}(2001)\citenamefont {Jones},
  \citenamefont {Oliphant}, \citenamefont {Peterson} \emph {et~al.}}]{numpy}%
  \BibitemOpen
  \bibfield  {author} {\bibinfo {author} {\bibfnamefont {E.}~\bibnamefont
  {Jones}}, \bibinfo {author} {\bibfnamefont {T.}~\bibnamefont {Oliphant}},
  \bibinfo {author} {\bibfnamefont {P.}~\bibnamefont {Peterson}},  \emph
  {et~al.},\ }\href {http://www.scipy.org/} {\enquote {\bibinfo {title}
  {{SciPy}: Open source scientific tools for {Python}},}\ } (\bibinfo {year}
  {2001})\BibitemShut {NoStop}%
\bibitem [{\citenamefont {Huang}()}]{tensorboardX}%
  \BibitemOpen
  \bibfield  {author} {\bibinfo {author} {\bibfnamefont {T.-W.}\ \bibnamefont
  {Huang}},\ }\href@noop {} {\enquote {\bibinfo {title} {tensorboard{X}},}\
  }\bibinfo {note} {\url{https://github.com/lanpa/tensorboardX}}\BibitemShut
  {NoStop}%
\bibitem [{\citenamefont {Collette}(2013)}]{h5py}%
  \BibitemOpen
  \bibfield  {author} {\bibinfo {author} {\bibfnamefont {A.}~\bibnamefont
  {Collette}},\ }\href@noop {} {\emph {\bibinfo {title} {Python and HDF5}}}\
  (\bibinfo  {publisher} {O'Reilly},\ \bibinfo {year} {2013})\BibitemShut
  {NoStop}%
\bibitem [{Note1()}]{Note1}%
  \BibitemOpen
  \bibinfo {note} {To install run the following command: pip install
  schnetpack}\BibitemShut {NoStop}%
\bibitem [{Note2()}]{Note2}%
  \BibitemOpen
  \bibinfo {note} {Code can be found here:
  https://github.com/atomistic-machine-learning/schnetpack}\BibitemShut
  {NoStop}%
\bibitem [{\citenamefont {Prechelt}(1998)}]{prechelt1998early}%
  \BibitemOpen
  \bibfield  {author} {\bibinfo {author} {\bibfnamefont {L.}~\bibnamefont
  {Prechelt}},\ }in\ \href@noop {} {\emph {\bibinfo {booktitle} {Neural
  Networks: Tricks of the trade}}}\ (\bibinfo  {publisher} {Springer},\
  \bibinfo {year} {1998})\ pp.\ \bibinfo {pages} {55--69}\BibitemShut {NoStop}%
\end{thebibliography}%

\clearpage
\appendix

\section{Details on Experiments}\label{app:experiments}

\begin{figure}[htb]
\centering
\includegraphics[width=0.5\textwidth]{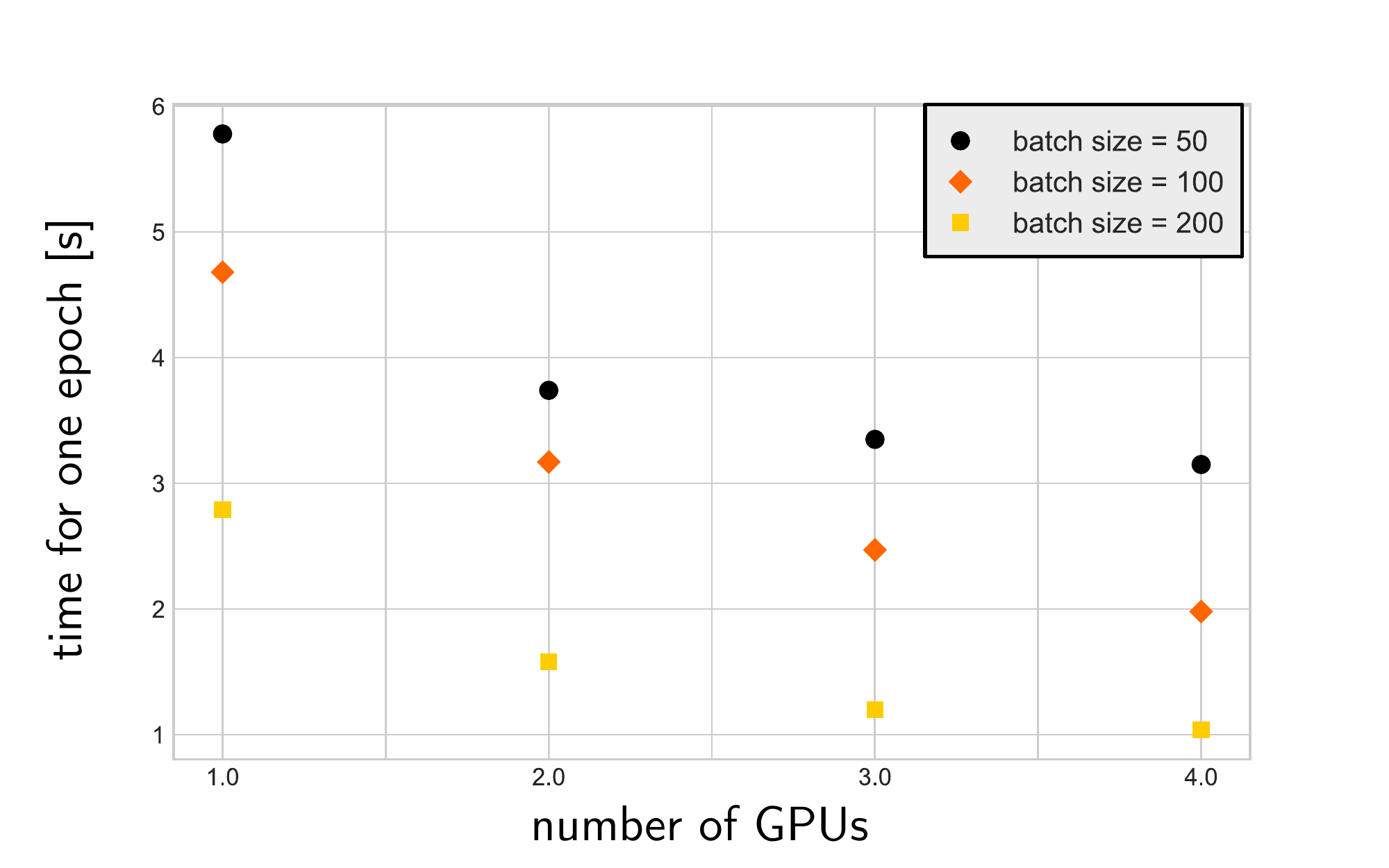}
\caption{Time required to go through one training epoch using 110k QM9 molecules for training.}
\label{fig:gpuscaling}
\end{figure}

We train three models on different splits which are summarized in Table~\ref{tab:detailsexp}. The reported errors and runtimes are their average. We use a Tesla P100 GPU for training. The scripts which include all the choices for the hyperparameters are contained in the scripts subdirectory of SchNetPack. One of the main reasons why we can train SchNet models faster than in the original publications~\cite{schutt2017schnet, schutt2018schnet} is that we are using a different learning rate schedule. In the original publications, the learning rate was lowered by a factor of 0.96 every 100k iterations. In our experiments however, the learning rate is reduced by a decay factor if the validation loss has not improved over a given number of epochs. We refer to Tables \ref{tab:hyperdetailsSchNet} to \ref{tab:hyperdetailsACSFs} for details on learning rates, schedules and symmetry function compositions. While both learning rate schedules lead to comparable results in the long run, our new setup converges significantly faster.

We can further speed up training by taking advantage of the support for multiple GPUs in SchNetPack as is demonstrated in Figure~\ref{fig:gpuscaling}. In our experiments, we only did so for the ANI-1 dataset.

\begin{figure*}[h]
\centering
\includegraphics[width=0.8\textwidth]{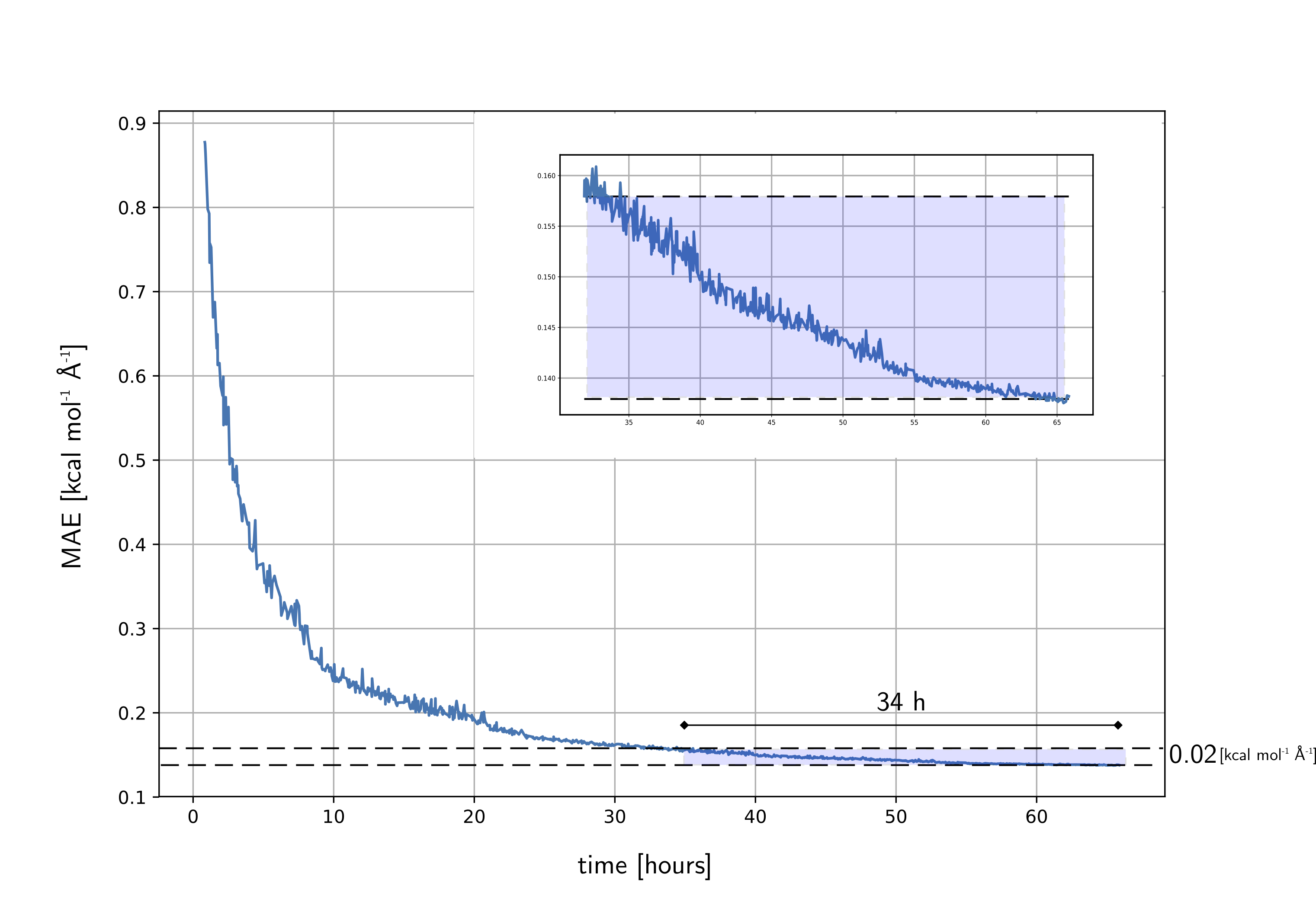}
\caption{Force training of SchNet on acetylsalicylic acid (N=50k). Note that approximately 50\% of training time is spend on fine-tuning the last $\sim0.02$ kcal mol$^{-1}${\AA}$^{-1}$.}
\label{fig:lrcurves}
\end{figure*}

Finding the optimal point to stop the training process is a well-known problem in the optimization of neural networks~\cite{prechelt1998early}.
It is worth noting that a significant amount of the training time in our experiments is spent on fine-tuning the accuracy of the prediction.
Figure~\ref{fig:lrcurves} demonstrates this for the example of training SchNet on 50k reference calculations for acetylsalicylic acid. In this case, approximately 50\% of the training time is spend on the last 0.02 kcal/mol improvement.
This also means that if this minor improvement is not required, the training can already be stopped after half of the reported training time when the learning progress flattens out.
We plan to implement more advanced learning rate schedules and stopping criteria in future versions to alleviate this issue.

\begin{table*}[h]
\caption{Setup of experiments \label{tab:detailsexp}}
\centering
\begin{ruledtabular}
\begin{tabular}{lrrr} 
Dataset & Train & Validation & Test \\
\colrule
Malondialdehyde (N=1k) & 950 & 50 & 992,237 \\ 
\colrule
Malondialdehyde (N=50k) & 49,000 & 1,000 & 943,237 \\ 
\colrule
Acetylsalicylic acid (N=1k) & 950 & 50 & 210,762 \\ 
\colrule
Acetylsalicylic acid (N=50k) & 49,000 & 1,000 & 161,762 \\ 
\colrule
QM9 & 109,000 & 1,000 & 20,813 \\ 
\colrule
ANI-1 (N=10.1M) & 10,000,000 & 100,000 & 11,957,374 \\ 
\colrule
ANI-1 (N=19.8M) & 17,600,000 & 2,200,000 & 2,257,374 \\ 
\colrule
Materials Project & 60,000 & 2,000 & 21,623 \\ 
\end{tabular}
\end{ruledtabular}
\end{table*}

\begin{table*}[h]
\caption{Setup for SchNet training \label{tab:hyperdetailsSchNet}}
\centering
\begin{ruledtabular}
\begin{tabular}{lllcrrr} 
Dataset & Learning Rate & Decay Factor & Minimal Learning Rate & Patience & $\rho$ & Batch Size \\
\colrule
QM9 & 0.0001 & 0.5 &1e-06 & 25 & -- & 100 \\ 
\colrule
ANI-1 & 0.0001 & 0.5 &1e-06 & 6 & -- & 400 \\ 
\colrule
MD17 (1k) & 0.0001 & 0.5 &1e-06 & 150 & 0.1 & 100 \\ 
\colrule
MD17 (50k) & 0.0001 & 0.5 &1e-06 & 50 & 0.1 & 100 \\ 
\colrule
Materials Project & 0.001 & 0.5 &1e-06 & 25 & -- & 32 \\ 
\end{tabular}
\end{ruledtabular}
\end{table*}

\begin{table*}[h]
\caption{Setup for Behler--Parrinello training \label{tab:hyperdetailsBehler}}
\centering
\begin{ruledtabular}
\begin{tabular}{lllcrrr} 
Dataset & Learning Rate & Decay Factor & Minimal Learning Rate & Patience & $\rho$ & Batch Size \\
\colrule
QM9 & 0.001 & 0.5 &1e-06 & 25 & - & 100 \\ 
\colrule
MD17 (1k) & 0.01 & 0.5 &1e-06 & 20 & 0.1 & 20 \\ 
\colrule
MD17 (50k) & 0.01 & 0.5 &1e-06 & 20 & 0.1 & 100 \\ 
\end{tabular}
\end{ruledtabular}
\end{table*}

\begin{table*}[h]
\caption{Setup for ACSFs and wACSFs used in the Behler--Parrinello models. The symmetry functions were standardized in all experiments.\label{tab:hyperdetailsACSFs}}
\centering
\begin{ruledtabular}
\begin{tabular}{ccrrrc} 
Dataset & Type & $n_\mathrm{rad}$ & $n_\mathrm{ang}$ & $n_\mathrm{SF}$ & centered radial \\
\colrule
\multirow{2}{*}{QM9} &  ACSF &  5 & 3 & 115 & - \\
                        & wACSF & 22 & 5 & 32 & - \\ 
\colrule
\multirow{2}{*}{MD17 (1k)} &  ACSF &  5 & 3 &  51 & + \\
                           & wACSF & 15 & 18 & 51 & + \\ 
\colrule
\multirow{2}{*}{MD17 (50k)} &  ACSF &  5 & 3 &  51 & + \\
                            & wACSF & 15 & 18 & 51 & + \\
\end{tabular}
\end{ruledtabular}
\end{table*}


\end{document}